\documentclass[aps,prb,twocolumn,superscriptaddress,showpacs,english]{revtex4-1}

\usepackage[T1]{fontenc}
\usepackage[latin9]{inputenc}
\usepackage{babel}
\usepackage{amsmath}
\usepackage{amssymb}
\usepackage{wasysym}
\usepackage{graphicx}
\usepackage{xcolor}
\usepackage{graphicx}

\usepackage[linktocpage=true,
  colorlinks=true, 
  pdfborder={0 0 0},
  linkcolor=blue,
  citecolor=blue,
  filecolor=yellow,
  urlcolor=blue,
  bookmarks,
  pdfauthor={},
]{hyperref}

\newcommand{\Basel}{Department of Physics, Universit\"at Basel, Klingelbergstr. 82, 4056 Basel, Switzerland}
\newcommand{\Halle}{Max-Planck Institut f\"ur Microstrukture Physics, Weinberg 2, 06120 Halle, Germany}
\newcommand{\HalleUni}{Institut f\"ur Physik, Martin-Luther-Universit\"at Halle-Wittenberg, D-06099 Halle, Germany }

\newcommand{\tc}{T$_{\textmd C }$}
\newcommand{\ho}{H$_{\rm 2}$O}
\newcommand{\sh}{H$_{\rm 3}$S}

\newcommand{\omlog}{$\omega_{\textmd log}$}

\begin{document}

\title{Emergence of superconductivity in doped H$_2$O ice at high pressure}

\author{Jos\'e~A. Flores-Livas} \affiliation{\Basel} 
\author{Antonio Sanna}          \affiliation{\Halle}
\author{Arkadiy Davydov}         \affiliation{\Halle}
\author{Stefan Goedecker}       \affiliation{\Basel}
\author{Miguel~A.~L. Marques}   \affiliation{\HalleUni}

\date{\today}

\begin{abstract}
We investigate the possibility of achieving high-temperature
superconductivity in hydrides under pressure by inducing metallization of 
otherwise insulating phases through  doping, a path previously used to render 
standard semiconductors superconducting at ambient pressure. 
Following this idea, we study \ho, one of the most abundant and well-studied substances, 
we identify nitrogen as the most likely and promising substitution/dopant. 
We show that for realistic levels of doping of a few percent, the phase X 
of ice becomes superconducting with a critical temperature of about 60~K at 150~GPa. 
In view of the vast number of hydrides that are strongly covalent bonded, 
but that remain insulating until rather large pressures, our results open a series of
new possibilities in the quest for novel high-temperature superconductors. 
\end{abstract}

\pacs{~}
\maketitle

\section{introduction}

The theoretical prediction~\cite{Duan_SciRep2014} and subsequent 
experimental discovery~\cite{DrozdovEremets_Nature2015} of
superconductivity in \sh\ at 200~GPa, with the record critical
temperature (\tc) of 203~K, rekindled the century-old dream of a room
temperature superconductor. The mechanism for superconductivity is
clearly understood within the strong coupling theory of Bardeen,
Cooper, and Schriffer~\cite{Theory_of_superconductivity} and the high \tc\ arises from the strong
electron-phonon coupling due to the peculiar electronic structure of 
this system under 
pressure~\cite{mazin2015superconductivity,Maramatsu-Hemley_2015,SH_PRB-Mazin-2015,PRB_Duan2015,Heil-Boeri_PRB2015,Flores-Livas_H3Se2016}. 
The aim of this research effort is to better understand how high 
critical temperatures can be achieved and if the same mechanisms can
work at lower pressures and/or even higher (room temperature) \tc\ in
other materials. In this quest for novel high-\tc\ superconductors,
many other materials have been proposed. As the presence of hydrogen
seems to be fundamental to reach the very high phonon frequencies,
strong electron-phonon coupling, and therefore large critical
temperatures~\cite{Ashcroft_PRL1968,RichardsonAshcroft_PRL97,Cudazzo_PRL2008,mcmahon_high_2011,PRB_H-tc_erratum},
the major emphasis has been given to other
hydrides~\cite{Ashcroft_PRL2004,tse_novel_2007,Chen_PNAS2008,Kim_PNAS2008,FengAsHoffman_Nature2008,Wang_PNAS2009,Yao_PNAS2010,gao_high-pressure_2010,Kim_PNAS2010,Li_PNAS2010,Zhou_PRB2012,Hooper_JPC-2014,Maramatsu-Hemley_2015,esfahani2016superconductivity,struzhkin2014synthesis,PRL_TeH3_2016} such as
silane~\cite{Eremets_Science2008,Chen_PNAS2008,degtyareva_formation_2009,Hanfland_PRL2011,Strobel_PRL2011},
disilane~\cite{Disilane_JAFL}, 
hydrogen sulfide~\cite{errea2016quantum,SH_PRB-Mazin-2015,Errea_anhaPRL2015,PRB_Duan2015,PRB_akashi_2015_HS,Heil-Boeri_PRB2015,quan_impact_2015,xie2014superconductivity,ortenzi_TB_2015,akashi_mangeli-phases} 
hydrogen selenide~\cite{Flores-Livas_H3Se2016}, phosphine~\cite{Drozdov_ph3_arxiv2015,ours_PH_rPRB2016}, etc.

Unfortunately, many (if not most) chemical compounds containing
hydrogen only metallize at extremely high pressures. The paradigmatic
case is pure hydrogen, whose metallic state is the ground-state structure only above
500~GPa~\cite{Wigner_JCP1935,pickard_structure_2007,LeToullec2002,Eremets_NatMat2011,Hemley_PRL2012,HRussell_hydrogenJACS2014}. There
are certainly other phases that are metallic at lower pressure, but
these are often thermodynamically unstable, and therefore difficult,
if not impossible, to access experimentally.

A possible, but until now overlooked, solution is doping. It is well
known that by introducing enough electron- or hole-donating impurities
one can render a semiconducting system metallic and even
superconducting. This strategy was already successful in inducing
superconductivity in diamond (doped with boron) in
2004~\cite{ekimov_superconductivity_2004}, silicon (doped with
boron~\cite{bustarret_superconductivity_2006}), germanium (doped with
gallium~\cite{herrmannsdorfer_superconducting_2009}), and silicon
carbide (doped with boron~\cite{kriener_superconductivity_2008} or
aluminum~\cite{muranaka_superconductivity_2008}). Transition
temperatures are unfortunately rather low, remaining below 4~K.

In this work we follow this strategy, and investigate if the
combination of doping and high pressure can be used to obtain
high-temperature superconductivity in hydrides.   
We select as an example one of the most abundant, and also one of the best studied,
hydrides, namely \ho. Note that undoped H$_2$O remains insulating up to
the terapascal range of pressures. In fact, its metallization was
predicted to occur beyond
5~TPa~\cite{militzer2010_prl-ice_cmcm,mcmahon2011_rPRB_Ice-tera,H4O_coreplanet_PRB2013,H2O2_Pickard-Needs_PRL2013}. 
The paper is organized as follows: 
in Sec.II, we present the discussion of the structural transition from molecular ice to the covalent phase under pressure, 
followed, in  Sec.III,  by  the  results  for  the  electronic  structure of doped ice.  
In  Sec.IV,  we  discuss  the emergence of superconductivity in ice under pressure. 
Finally, we present our conclusions in Sec. V.

\section{Covalent phase of Ice under pressure}

\begin{figure}[t]
\includegraphics[width=1.0\columnwidth,angle=0]{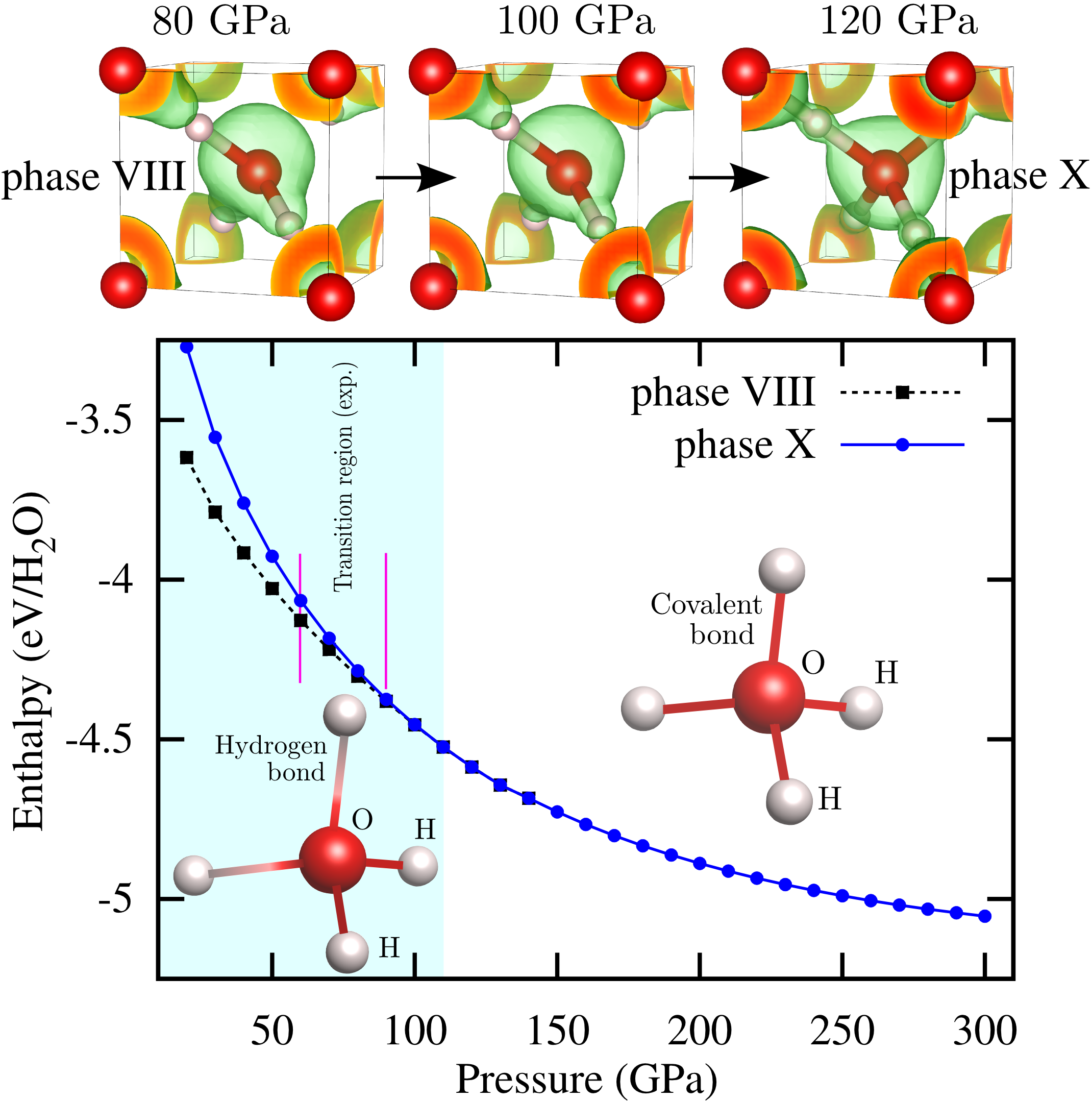}
\caption{(Color online) Top panel: Calculated electron localization
  function (ELF) at $n=$0.6 for ice as a function of pressure. 
  The molecular crystal transforms to a fully covalent phase-X at
  120~GPa. Bottom panel: Calculated enthalpy for the VIII and X phases
  of H$_2$O ice.  Values are given with respect to the elemental
  decomposition H$_2$ + O. The reference structures for hydrogen are
  $P6_3m$ (0--120~GPa) and $C2/c$ (120--300~GPa) from
  Ref.~\cite{pickard_structure_2007}.  
  The reference phase for oxygen is $C2/m$, $\zeta$ structure~\cite{Ma-Oganov_oxygen_PRB2007,Ochoa-calle_oxygen-hibrids_PRB2015,Mezouar_c2m-phase-Oxyg_PRL2009}. 
  The experimental transition region between the two
  structures is marked with lines according to Goncharov et al.~\cite{Goncharov_PRL_Raman1998}. 
  From our calculations, the transition from a molecular ice-crystal to the fully symmetric covalent phase X occurs at 120\,GPa.}
\label{fig:enthalpy}
\end{figure}

Despite its simple chemical formula, \ho\ appears in nature in all
three common states of matter and it has one of the most complex phase
diagram known~\cite{REVIEW_2012ice}.  Over a dozen of different
crystallographic phases have been reported or predicted in a wide range
of temperature and pressures~\cite{bjerrum1952structure,kamb1964ice,londono1993neutron,whalley1968ice,kuhs1984structure,
kamb1967structure,salzmann2006preparation,kuhs1984structure,jorgensen1985disordered,besson1994variation,hemley1987static,goncharov2005dynamic,
PRL_Vos-H2_Clathrate1993,mishima1994reversible,loerting2001second,smith1999existence,angell2004amorphous,yen2015dielectric,
militzer2010_prl-ice_cmcm,mcmahon2011_rPRB_Ice-tera,H4O_coreplanet_PRB2013,H2O2_Pickard-Needs_PRL2013}.

At ambient pressure and low temperatures ice
assumes~\cite{bjerrum1952structure} its phase I, where oxygen has four
hydrogen neighbors: two covalently bonded (forming the \ho\ molecule)
and two connected by hydrogen bonds to neighboring H$_2$O molecules.
Below 200~K, phase I transforms to phase XI, and under compression to
phase IX, stable in the range from 0.1 to 1\,GPa. Under further
compression, and at very low temperatures, the phase VIII dominates up
to 60--80~GPa. This molecular crystal can be seen as an ordered and
symmetrized version of phase VII that occurs at high temperatures. 
At 80--90\,GPa was reported the emergence of the cuprite-type ice-X, 
characterized by static, symmetric O--H bonds~\cite{goncharov1996compression,Goncharov_PRL_Raman1998,bove2015effect}. 

Figure~\ref{fig:enthalpy} shows our theoretical phase diagram under
pressure.  The energies, atomic forces and stresses necessary to
construct this figure were evaluated within density functional theory
with the Perdew-Burke-Erzernhof (PBE)~\cite{PBE96} approximation to
the exchange-correlation functional. A plane wave basis-set with a
high cutoff energy of 1000~eV was used to expand the wave-functions
together with the projector augmented wave (PAW) method as implemented
in the Vienna Ab Initio Simulation Package~{\sc vasp}~\cite{VASP_Kresse}. 
Geometry relaxations were performed with tight convergence criteria such 
that the forces on the atoms were less than 2~meV/\AA\ and the stresses were less than 0.1~meV/\AA$^3$.

At low pressure between 20 to 110~GPa we have phase VIII (see
Fig.~\ref{fig:enthalpy}, in agreement with the experimental phase diagram). 
Above 110~GPa it undergoes a transition to the proton-symmetric and experimentally confirmed phase X. 
It has been shown that due to proton symmetrization, quantum effects and anharmonicity no longer play a major role~\cite{He-H2O_ice_PRB2016,Hermann_H2O_ZPE_PRB_2013,benoit1998tunnelling} at higher pressure. 
Despite the pressure shift, our calculations (Fig.~\ref{fig:enthalpy}) are in good agreement with experiments.   
Phase X is the dominant structure of ice up to at least 300~GPa.  
This phase is extremely interesting from our point of view, as it is no longer a molecular crystal and exhibits a
complete covalent character, as indicated by the behavior of the 
electron localization function~\cite{Becke_ELF_JChPh1990} (see top panel of Fig.~\ref{fig:enthalpy}). 
This is absolutely essential for the appearance of doping induced superconductivity, otherwise 
impurities would just introduce localized states that can not participate in the formation of Cooper pairs.

\begin{figure*}[htb]
  \begin{center}
    \includegraphics[width=2.0\columnwidth,angle=0]{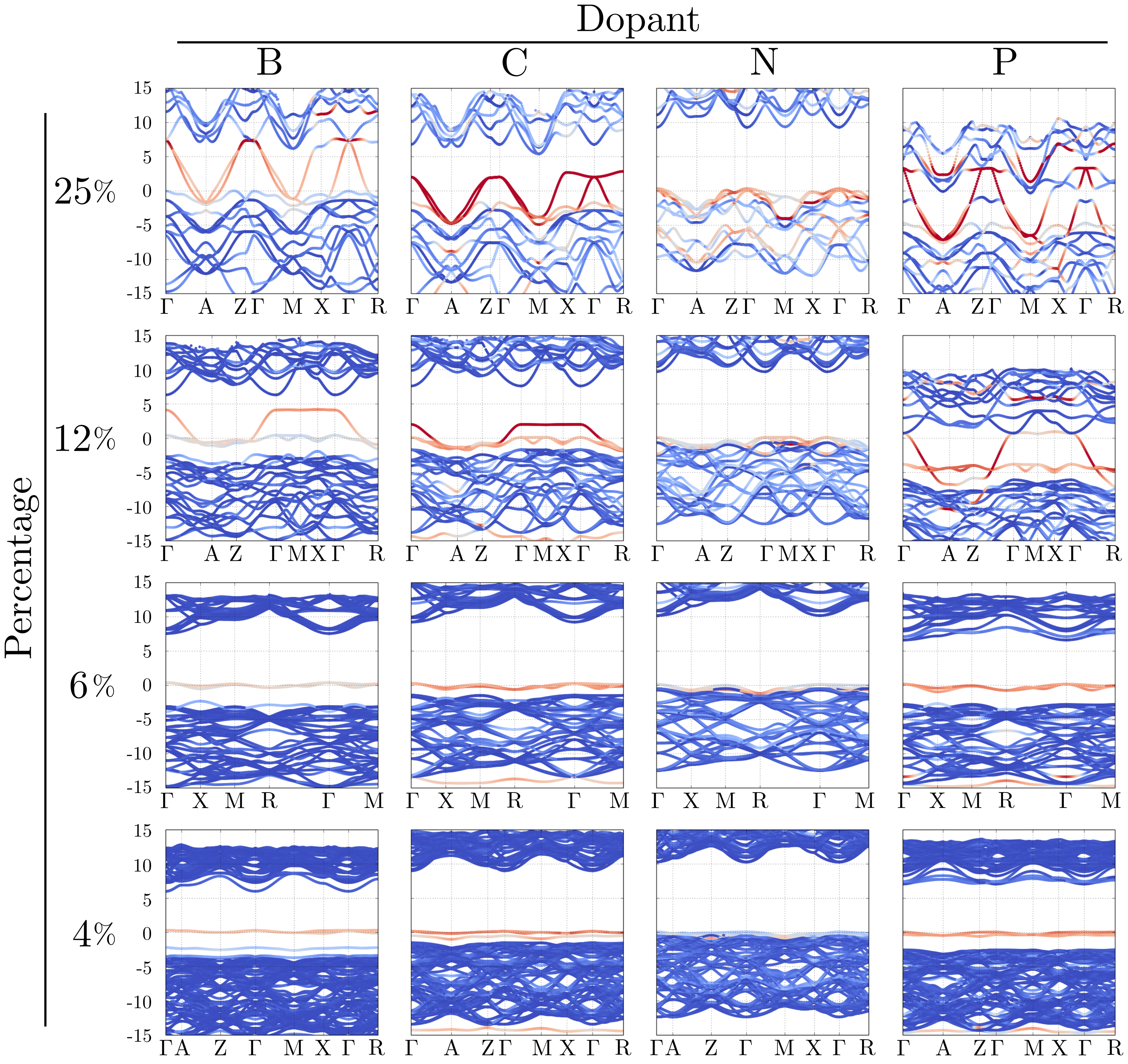}
    \caption{(Color online) Electronic band structure for different
      dopants (B, C, N and P) in the phase-X of ice at 150\,GPa. 
      The Fermi level is set to 0\,eV and the color scale represents the overlap of the Kohn-Sham states on the
      atomic orbitals of the dopant: red means large overlap (dopant states projected), gray intermediate, 
      while blue means small (hydrogen and oxygen states projected).}
    \label{fig:dopants_EDOS}
  \end{center}
 \end{figure*}
 
\section{Electronic structure of doped ice under pressure}

In order to study a realistic doping in ice, we have created supercell structures of ice-X under pressure for 
a wide range of doping values H$_2$O$_{1-x}$Dopant$_x$, with $x=25$\%, $12.5$\%,
$6.25$\%, and $4.16$\% ($x=100$\% indicates the removal of one electron per formula unit of \ho).  
Full structural relaxation were then carried out for the supercells (12 atoms cell for $x=25$\%, 
24 atoms cell for $12.5$\%, 48 atoms cell for $6.25$\% and 72 atoms cell for $4.16$\% ). 
For low doping (4--6\%), we only find fairly small modifications of the
crystal structure of ice-X.  On the contrary, larger doping levels 
lead to a considerable deformation of the local environment.

Figure~\ref{fig:dopants_EDOS} depicts the electronic band structure
obtained for boron, carbon, nitrogen and phosphorous used as dopant in the ice-X at 150\,GPa. 
In these plots the color scale represents the overlap of the Kohn-Sham states on the
atomic orbitals of the dopant: red means large overlap (dopant states projected), gray intermediate, while blue means small (hydrogen and oxygen states projected). 
For large doping with boron and phosphorous (12\% and 25\%), the
dispersive band coming from the dopant completely closes the gap in ice-X, 
while for 4\% and 6\% the dopant states form impurity molecular-like bands. 
These atoms are therefore not suitable to hole-dope ice-X. The case of carbon is intermediate: at low doping we
see again the formation of impurity bands, while at higher levels we do see some hybridization 
between the carbon and the ice bands at the top of the valence. I
n general we find a higher density of states at the Fermi level with B, C and P acting as a dopant, 
however the band structure shows mostly localized 
molecular states which are detrimental to superconductivity. 
It is important to mention that, since the ice-X is highly symmetric
(cubic structure, $P-43m$, space group 215) there is only one site
symmetry to substitute for high doping levels (i.e. 25 and 12 percent) and for a reasonably small supercell. 
For lower doping levels, however, many other doping sites become available. 
Nevertheless, a study of all possible site/defect substitutions of oxygen by dopants is clearly beyond the scope 
of this work (and would probably require techniques such as cluster expansion). 
However, our results provide a clear general trend of the physics of doped ice under pressure.

\begin{figure}[htb]
\includegraphics[width=0.9\columnwidth,angle=0]{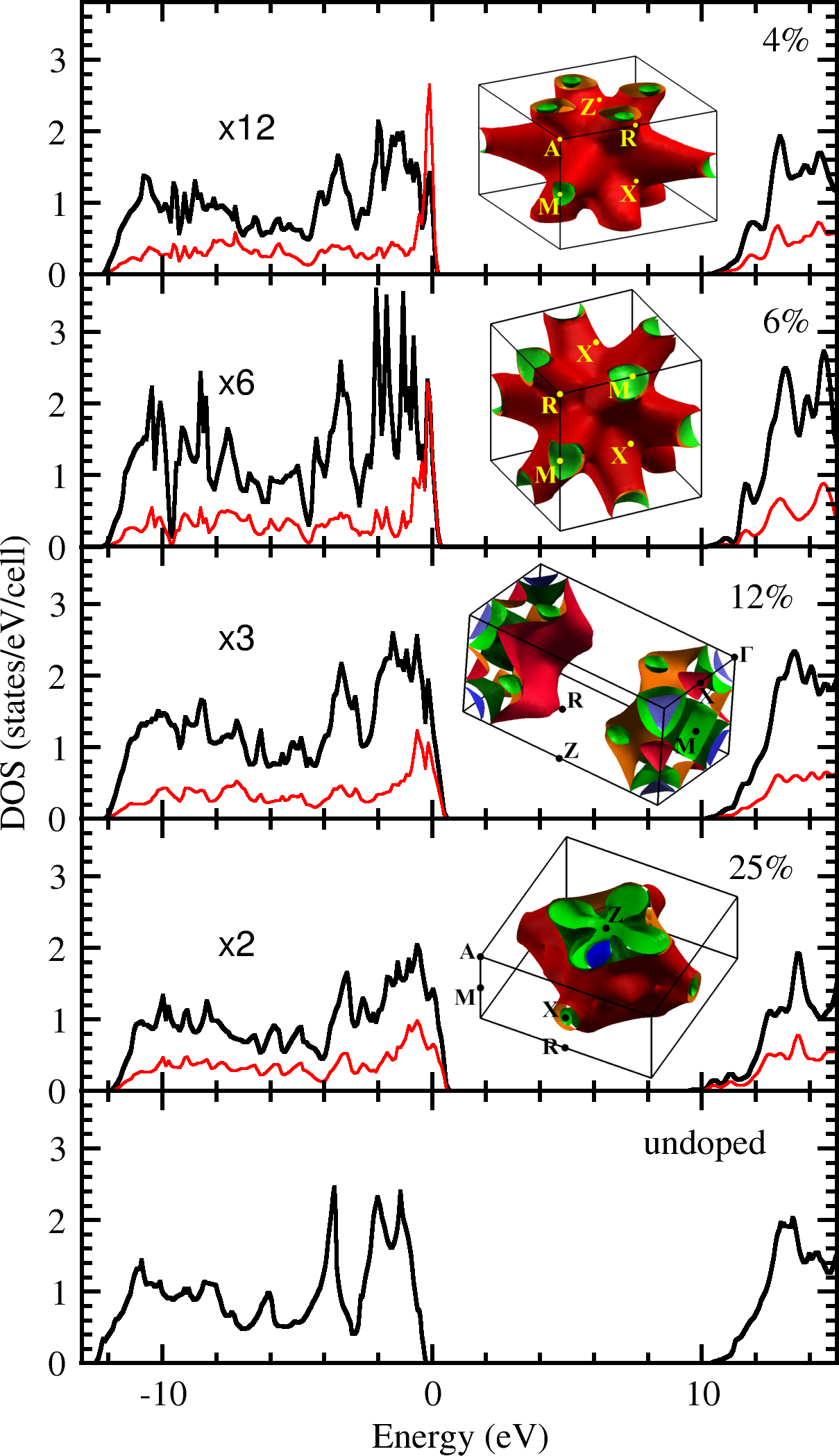}
\caption{(Color online) Density of states (DOS) and Fermi surfaces
  of undoped and N-doped phase X of \ho\ at 150\,GPa.  Thick black
  lines are the total DOS (a scaling factor is applied for plotting
  convenience). Solid red curves are the projection on nitrogen atomic
  states. All doped systems are metallic featuring one or two small
  electron pockets around the gamma point (blue/green) an one or two
  large open surfaces (red/green). }
\label{fig:dos_fs}
\end{figure}

Conversely, to what the other elements have shown, nitrogen clearly induces hole doping in ice-X (seeing in Figure~\ref{fig:dopants_EDOS}) and, as shown in Ref.~\onlinecite{Henning-Sanna-Marques_pettifor} (supplemental) it is the {\it statistically} most likely non iso-valent element able to substitute oxygen. 
The bottom panel in Fig.~\ref{fig:dos_fs} shows in detail the electronic structure of ice-X.  
Phase-X is an insulator with a PBE electronic band gap about 10\,eV and does not undergo major 
modifications at least up to 300\,GPa. 
According to our calculations, the insulator-metal transition is achievable starting from 
values of 4\% nitrogen doping (as seen in the projected-states band plot Figure~\ref{fig:dopants_EDOS}), 
we can conclude that holes are indeed populating the Fermi level. 
For 25\% doping, the top valence is mostly dominated by nitrogen states forming a fairly dispersive band. 
For values between 4\% and 12\%, oxygen and nitrogen strongly hybridize in the top valence forming 
the metallic states (gray colors in the plot).

Electronically, it is clear that nitrogen doping introduces holes in
the ice-X crystal. These hole states are, as intended, hybrid O--N
states, as can be seen from the projected density of states in
Fig.~\ref{fig:dos_fs}. At high doping ($12.5$\% and $25$\%) the states at
the Fermi level are homogeneous O--N hybrids, meaning that the density
of N states is simply proportional to the fraction of N atoms.  On the
other hand, at lower doping ($4.16$\% and $6.25$\%) the N projected DOS,
although overall smaller, is larger than the N/O fraction and shows a
sharp peak close to the Fermi level, indicating that the induced holes
are more localized on the N sites. We note that the two last values are realistic doping values 
and similar to the doping values used to render diamond or silicon superconducting at ambient pressure~\cite{blase_superconductivity_2011}.

\section{Emergence of Superconductivity in N-doped ice}

\begin{figure}[htb]
\includegraphics[width=1.0\columnwidth,angle=0]{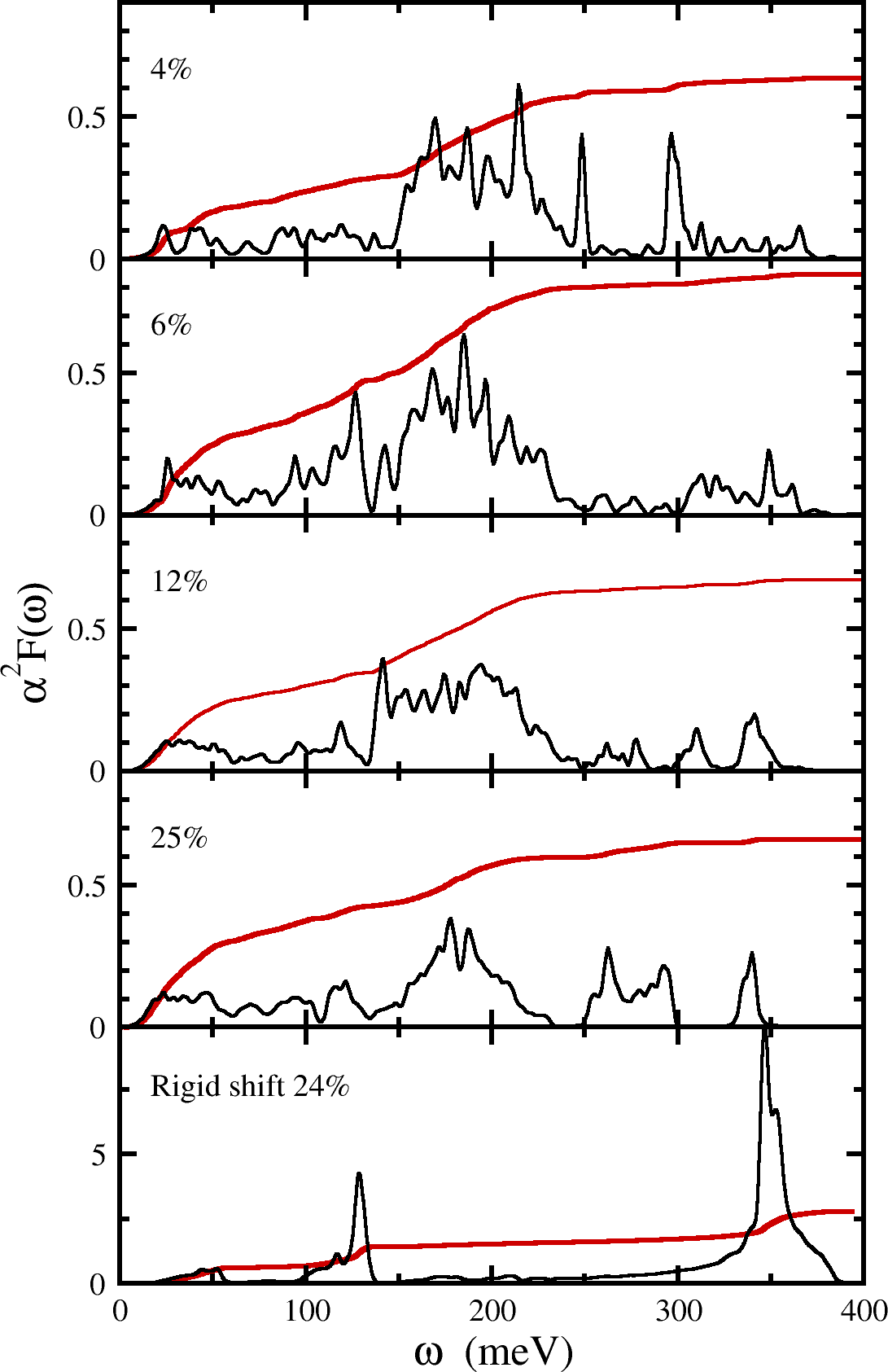}
\caption{(Color online) Eliashberg spectral function (black lines) and
  integration curve of the electron-phonon coupling constant
  $\lambda(\omega)$ (red lines) for hole-doped \ho\ in its phase X at
  150\,GPa. Doping level is indicated in each panel.}\label{fig:a2f}
\end{figure}

There are many possible ways to study theoretically the effect of doping in the superconducting properties. 
The simplest way is by shifting rigidly the Fermi level, leaving both Kohn-Sham eigenvalues and eigenfunctions unchanged. 
The resulting phonon spectrum and electron-phonon scattering amplitude can then be used within an Eliashberg~\cite{Eliashberg,AllenMitrovic1983}
scheme to compute the superconducting critical temperature as a function of the position of the Fermi level. 
For the ice-X of \ho\ at 150\,GPa (2 formula unit cell) we compute, within this procedure, an astonishingly high
 phononic superconducting coupling, leading to room-temperature superconductivity already at a doping of a few percent! 
Although widely used in the literature, we can not expect that such a rigid shift of the Fermi level yields more than an estimate for the order of 
magnitude of the critical temperature upon doping. 
In fact, the extreme electron-phonon coupling obtained by the rigid shift
 would induce a strong electronic response, leading to a complete breakdown of the rigid shift
approximation. Moreover, this method does not account for important 
 physical effects stemming from the metallic part of the electronic screening, such as the mechanism responsible for Kohn anomalies~\cite{KohnAnomaly} 
 that can significantly modify the spectrum of phonons.

 \begin{figure}[htb]
\includegraphics[width=1.0\columnwidth,angle=0]{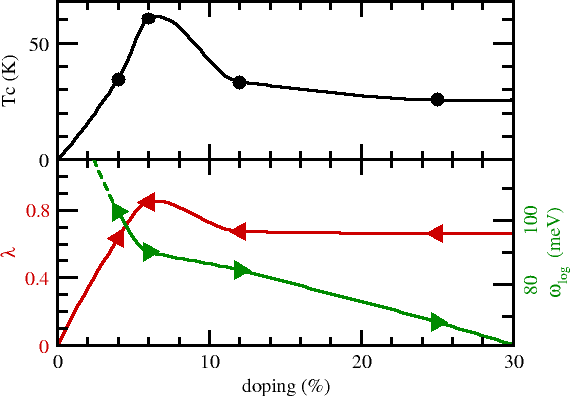}
\caption{(Color online) Top panel: calculated critical temperatures
  with the McMillan-Allen-Dynes formula as a function of doping of
  H$_2$O in its phase X at 150~GPa. Lower panel: electron phonon
  coupling constant $\lambda$ (red left triangles and left axis) and 
  average phonon frequency \omlog (green right triangles and right axis). 
  Solid lines are a guide to the eye.}
\label{fig:tc}
\end{figure}

A more realistic way to study theoretically the effect of doping in the superconducting properties is 
to calculate the phonon and electronic-phonon matrix elements~\footnote{The phonon spectra and the electron-phonon matrix
elements were obtained employing density-functional perturbation
theory~\cite{Baroni_1987a,2n+1_Gonzepaper}, as implemented in the
plane-wave based code {\sc abinit}~\cite{gonze_abinit_2009}. For the
electron-phonon the following $k$ and $q$-meshes were used for the
different supercells: 25\% doping, $k=16\times16\times16$, and
$q=8\times8\times8$; 12.5\% doping, $k=8\times8\times8$ and
$q=4\times4\times4$; 6.13\% doping, $k=4\times4\times4$ and
$q=4\times4\times4$; 4.17\% doping, $k=2\times2\times2$ and
$q=2\times2\times2$.} in the supercell of nitrogen doped ice-X.  
As we have seen, doping turns out to have a dramatic effect in the electronic structure of ice, 
and in the phonon spectrum (not shown) and in the Eliashberg spectral functions, shown in Fig.~\ref{fig:a2f} as a function of doping at 150\,GPa. 

Comparing the supercell calculations with results obtained with a
rigidly shifted Fermi level (see Fig.~\ref{fig:a2f}), we observe a
complete restructuring of the phonon energies and coupling strength.
The metallization provides a significant screening both causing a
softening of the phonon frequencies and a reduction of the deformation
potential.  Nevertheless we still observe a significant
electron-phonon coupling, as can be seen from the Eliashberg functions
as well as from the logarithmic average of the phonon frequency
\omlog~\cite{Carbotte_RMP1990,AllenMitrovic1983} (see Fig.~\ref{fig:tc}). 
We furthermore calculated the phonon band-structure for B, C and P doping. 
We find that all systems are highly unstable with large imaginary frequencies, with the only exception nitrogen, being dynamically stable in the doped range studied. 

For low doping ($<12$\%) two major contributions to $\lambda$ can be
distinguished: i) the low frequency optical phonons (oxygen
vibrations) that couple with $2p$ nitrogen states; and ii) the
mid-frequency range 150--200 meV (1100--1800\,cm$^{-1}$) that couples
with the covalent oxygen-nitrogen hybridized state.  For 25\% doping
all phonon branches contribute significantly to the $e-p$ coupling,
since at this doping limit the structure adopts a completely metallic
character.

From these parameters we can easily estimate the critical temperature
by means of the McMillan-Allen-Dynes~\cite{AllenDynes_PRB1975} formula
(assuming $\mu^*$=0.1). This gives \tc\ in the range from 20 to 60~K,
with the maximum value reached for a doping level of 6.25\% (see
Fig.~\ref{fig:tc}). Although lower than the astonishing value found in
sulphur hydride (200\,K) and considerably lower than the rigid shift
prediction (300\,K) this is still a sizable value, much larger than the
\tc$\lesssim4$~K found in doped semiconductors at ambient pressure.

A possible path to reach the synthesis of the nitrogen doped ice-X and the superconducting state 
is to start from a high pressure synthesis similar to what is used to obtain  H$_2$+H$_2$O clathrates. 
In 1993, Vos et al.~\cite{PRL_Vos-H2_Clathrate1993} reported the formation of filled-ice clathrates. 
They succeed to fill, at room temperature, H$_2$ molecules inside H$_2$O crystalline $C_1$ (clathrate) 
phase at 0.7\,GPa, this structure is also viewed as ice-II phase.  
Experimentally, the unit cell of the $C_1$ phase contains 36 water molecules, in a channel-like arrangement, 
which can host up to six hydrogen molecules~\cite{Grigory_JPL_clathrate-2013,Quian_SR_hydrogen-clathrates2013}. 
In short we propose to start the synthesis from the analogous H$_2+$H$_2$O clatrhate ($C_1$) 
but filled with N$_2$ molecules where the percentage of filled N$_2$ molecules will determine the doping 
level at high pressure, and the percentage of filling of molecules can 
be controlled at ambient conditions~\cite{Strobel_H2_H2O_2011}. 

\section{conclusions}

In conclusion, we have investigated the possibility of inducing
high-temperature superconductivity by doping insulating hydrides under
pressure. By taking the phase X of ice as an example, we studied how
the phonon spectra and the electron-phonon coupling evolve as a function of nitrogen doping. 
Despite the number of elements and the range studied to dope H$_2$O it is clear that only nitrogen, favors thermodynamically the synthesis of the insulator-metal transition of ice under pressure.
It turns out that for rather reasonable values of doping, one can reach superconducting transition 
temperatures as high as 60\,K at 150\,GPa. Considering the vast number of hydrides that 
remain insulating under pressure and that can be doped, this result opens
a number of possibilities for the exploration of high-temperature superconductivity in these unique systems. 

\begin{acknowledgments}
J.A.F.-L. acknowledges computational resources under the project
(s499) from the Swiss National Supercomputing Center (CSCS) in Lugano. 
M.A.L.M. acknowledges partial support from the DFG through projects SFB-762 and MA 6787/1-1.
\end{acknowledgments}

\bibliographystyle{apsrev4-1}
\bibliography{ice_paper}

\begin{thebibliography}{102}%
\makeatletter
\providecommand \@ifxundefined [1]{%
 \@ifx{#1\undefined}
}%
\providecommand \@ifnum [1]{%
 \ifnum #1\expandafter \@firstoftwo
 \else \expandafter \@secondoftwo
 \fi
}%
\providecommand \@ifx [1]{%
 \ifx #1\expandafter \@firstoftwo
 \else \expandafter \@secondoftwo
 \fi
}%
\providecommand \natexlab [1]{#1}%
\providecommand \enquote  [1]{``#1''}%
\providecommand \bibnamefont  [1]{#1}%
\providecommand \bibfnamefont [1]{#1}%
\providecommand \citenamefont [1]{#1}%
\providecommand \href@noop [0]{\@secondoftwo}%
\providecommand \href [0]{\begingroup \@sanitize@url \@href}%
\providecommand \@href[1]{\@@startlink{#1}\@@href}%
\providecommand \@@href[1]{\endgroup#1\@@endlink}%
\providecommand \@sanitize@url [0]{\catcode `\\12\catcode `\$12\catcode
  `\&12\catcode `\#12\catcode `\^12\catcode `\_12\catcode `\%12\relax}%
\providecommand \@@startlink[1]{}%
\providecommand \@@endlink[0]{}%
\providecommand \url  [0]{\begingroup\@sanitize@url \@url }%
\providecommand \@url [1]{\endgroup\@href {#1}{\urlprefix }}%
\providecommand \urlprefix  [0]{URL }%
\providecommand \Eprint [0]{\href }%
\providecommand \doibase [0]{http://dx.doi.org/}%
\providecommand \selectlanguage [0]{\@gobble}%
\providecommand \bibinfo  [0]{\@secondoftwo}%
\providecommand \bibfield  [0]{\@secondoftwo}%
\providecommand \translation [1]{[#1]}%
\providecommand \BibitemOpen [0]{}%
\providecommand \bibitemStop [0]{}%
\providecommand \bibitemNoStop [0]{.\EOS\space}%
\providecommand \EOS [0]{\spacefactor3000\relax}%
\providecommand \BibitemShut  [1]{\csname bibitem#1\endcsname}%
\let\auto@bib@innerbib\@empty
\bibitem [{\citenamefont {Duan}\ \emph {et~al.}(2014)\citenamefont {Duan},
  \citenamefont {Liu}, \citenamefont {Tian}, \citenamefont {Li}, \citenamefont
  {Huang}, \citenamefont {Zhao}, \citenamefont {Yu}, \citenamefont {Liu},
  \citenamefont {Tian},\ and\ \citenamefont {Cui}}]{Duan_SciRep2014}%
  \BibitemOpen
  \bibfield  {author} {\bibinfo {author} {\bibfnamefont {D.}~\bibnamefont
  {Duan}}, \bibinfo {author} {\bibfnamefont {Y.}~\bibnamefont {Liu}}, \bibinfo
  {author} {\bibfnamefont {F.}~\bibnamefont {Tian}}, \bibinfo {author}
  {\bibfnamefont {D.}~\bibnamefont {Li}}, \bibinfo {author} {\bibfnamefont
  {X.}~\bibnamefont {Huang}}, \bibinfo {author} {\bibfnamefont
  {Z.}~\bibnamefont {Zhao}}, \bibinfo {author} {\bibfnamefont {H.}~\bibnamefont
  {Yu}}, \bibinfo {author} {\bibfnamefont {B.}~\bibnamefont {Liu}}, \bibinfo
  {author} {\bibfnamefont {W.}~\bibnamefont {Tian}}, \ and\ \bibinfo {author}
  {\bibfnamefont {T.}~\bibnamefont {Cui}},\ }\href@noop {} {\bibfield
  {journal} {\bibinfo  {journal} {Sci. Rep.}\ }\textbf {\bibinfo {volume}
  {4}},\ \bibinfo {pages} {6968} (\bibinfo {year} {2014})}\BibitemShut
  {NoStop}%
\bibitem [{\citenamefont {Drozdov}\ \emph
  {et~al.}(2015{\natexlab{a}})\citenamefont {Drozdov}, \citenamefont {Eremets},
  \citenamefont {Troyan}, \citenamefont {Ksenofontov},\ and\ \citenamefont
  {Shylin}}]{DrozdovEremets_Nature2015}%
  \BibitemOpen
  \bibfield  {author} {\bibinfo {author} {\bibfnamefont {A.~P.}\ \bibnamefont
  {Drozdov}}, \bibinfo {author} {\bibfnamefont {M.~I.}\ \bibnamefont
  {Eremets}}, \bibinfo {author} {\bibfnamefont {I.~A.}\ \bibnamefont {Troyan}},
  \bibinfo {author} {\bibfnamefont {V.}~\bibnamefont {Ksenofontov}}, \ and\
  \bibinfo {author} {\bibfnamefont {S.~I.}\ \bibnamefont {Shylin}},\ }\href
  {\doibase doi:10.1038/nature14964} {\bibfield  {journal} {\bibinfo  {journal}
  {Nature}\ }\textbf {\bibinfo {volume} {525}},\ \bibinfo {pages} {73}
  (\bibinfo {year} {2015}{\natexlab{a}})}\BibitemShut {NoStop}%
\bibitem [{\citenamefont {Bardeen}\ \emph {et~al.}(1957)\citenamefont
  {Bardeen}, \citenamefont {Cooper},\ and\ \citenamefont
  {Schrieffer}}]{Theory_of_superconductivity}%
  \BibitemOpen
  \bibfield  {author} {\bibinfo {author} {\bibfnamefont {J.}~\bibnamefont
  {Bardeen}}, \bibinfo {author} {\bibfnamefont {L.~N.}\ \bibnamefont {Cooper}},
  \ and\ \bibinfo {author} {\bibfnamefont {J.~R.}\ \bibnamefont {Schrieffer}},\
  }\href {\doibase 10.1103/PhysRev.108.1175} {\bibfield  {journal} {\bibinfo
  {journal} {Phys. Rev.}\ }\textbf {\bibinfo {volume} {108}},\ \bibinfo {pages}
  {1175} (\bibinfo {year} {1957})}\BibitemShut {NoStop}%
\bibitem [{\citenamefont {Mazin}(2015)}]{mazin2015superconductivity}%
  \BibitemOpen
  \bibfield  {author} {\bibinfo {author} {\bibfnamefont {I.~I.}\ \bibnamefont
  {Mazin}},\ }\href@noop {} {\bibfield  {journal} {\bibinfo  {journal}
  {Nature}\ }\textbf {\bibinfo {volume} {525}},\ \bibinfo {pages} {40}
  (\bibinfo {year} {2015})}\BibitemShut {NoStop}%
\bibitem [{\citenamefont {Muramatsu}\ \emph {et~al.}(2015)\citenamefont
  {Muramatsu}, \citenamefont {Wanene}, \citenamefont {Somayazulu},
  \citenamefont {Vinitsky}, \citenamefont {Chandra}, \citenamefont {Strobel},
  \citenamefont {Struzhkin},\ and\ \citenamefont
  {Hemley}}]{Maramatsu-Hemley_2015}%
  \BibitemOpen
  \bibfield  {author} {\bibinfo {author} {\bibfnamefont {T.}~\bibnamefont
  {Muramatsu}}, \bibinfo {author} {\bibfnamefont {W.~K.}\ \bibnamefont
  {Wanene}}, \bibinfo {author} {\bibfnamefont {M.}~\bibnamefont {Somayazulu}},
  \bibinfo {author} {\bibfnamefont {E.}~\bibnamefont {Vinitsky}}, \bibinfo
  {author} {\bibfnamefont {D.}~\bibnamefont {Chandra}}, \bibinfo {author}
  {\bibfnamefont {T.~A.}\ \bibnamefont {Strobel}}, \bibinfo {author}
  {\bibfnamefont {V.~V.}\ \bibnamefont {Struzhkin}}, \ and\ \bibinfo {author}
  {\bibfnamefont {R.~J.}\ \bibnamefont {Hemley}},\ }\href {\doibase
  10.1021/acs.jpcc.5b03709} {\bibfield  {journal} {\bibinfo  {journal} {J.
  Phys. Chem. C}\ }\textbf {\bibinfo {volume} {119}},\ \bibinfo {pages} {18007}
  (\bibinfo {year} {2015})}\BibitemShut {NoStop}%
\bibitem [{\citenamefont {Bernstein}\ \emph {et~al.}(2015)\citenamefont
  {Bernstein}, \citenamefont {Hellberg}, \citenamefont {Johannes},
  \citenamefont {Mazin},\ and\ \citenamefont {Mehl}}]{SH_PRB-Mazin-2015}%
  \BibitemOpen
  \bibfield  {author} {\bibinfo {author} {\bibfnamefont {N.}~\bibnamefont
  {Bernstein}}, \bibinfo {author} {\bibfnamefont {C.~S.}\ \bibnamefont
  {Hellberg}}, \bibinfo {author} {\bibfnamefont {M.~D.}\ \bibnamefont
  {Johannes}}, \bibinfo {author} {\bibfnamefont {I.~I.}\ \bibnamefont {Mazin}},
  \ and\ \bibinfo {author} {\bibfnamefont {M.~J.}\ \bibnamefont {Mehl}},\
  }\href {\doibase 10.1103/PhysRevB.91.060511} {\bibfield  {journal} {\bibinfo
  {journal} {Phys. Rev. B}\ }\textbf {\bibinfo {volume} {91}},\ \bibinfo
  {pages} {060511} (\bibinfo {year} {2015})}\BibitemShut {NoStop}%
\bibitem [{\citenamefont {Duan}\ \emph {et~al.}(2015)\citenamefont {Duan},
  \citenamefont {Huang}, \citenamefont {Tian}, \citenamefont {Li},
  \citenamefont {Yu}, \citenamefont {Liu}, \citenamefont {Ma}, \citenamefont
  {Liu},\ and\ \citenamefont {Cui}}]{PRB_Duan2015}%
  \BibitemOpen
  \bibfield  {author} {\bibinfo {author} {\bibfnamefont {D.}~\bibnamefont
  {Duan}}, \bibinfo {author} {\bibfnamefont {X.}~\bibnamefont {Huang}},
  \bibinfo {author} {\bibfnamefont {F.}~\bibnamefont {Tian}}, \bibinfo {author}
  {\bibfnamefont {D.}~\bibnamefont {Li}}, \bibinfo {author} {\bibfnamefont
  {H.}~\bibnamefont {Yu}}, \bibinfo {author} {\bibfnamefont {Y.}~\bibnamefont
  {Liu}}, \bibinfo {author} {\bibfnamefont {Y.}~\bibnamefont {Ma}}, \bibinfo
  {author} {\bibfnamefont {B.}~\bibnamefont {Liu}}, \ and\ \bibinfo {author}
  {\bibfnamefont {T.}~\bibnamefont {Cui}},\ }\href {\doibase
  10.1103/PhysRevB.91.180502} {\bibfield  {journal} {\bibinfo  {journal} {Phys.
  Rev. B}\ }\textbf {\bibinfo {volume} {91}},\ \bibinfo {pages} {180502}
  (\bibinfo {year} {2015})}\BibitemShut {NoStop}%
\bibitem [{\citenamefont {Heil}\ and\ \citenamefont
  {Boeri}(2015)}]{Heil-Boeri_PRB2015}%
  \BibitemOpen
  \bibfield  {author} {\bibinfo {author} {\bibfnamefont {C.}~\bibnamefont
  {Heil}}\ and\ \bibinfo {author} {\bibfnamefont {L.}~\bibnamefont {Boeri}},\
  }\href {\doibase 10.1103/PhysRevB.92.060508} {\bibfield  {journal} {\bibinfo
  {journal} {Phys. Rev. B}\ }\textbf {\bibinfo {volume} {92}},\ \bibinfo
  {pages} {060508} (\bibinfo {year} {2015})}\BibitemShut {NoStop}%
\bibitem [{\citenamefont {Flores-Livas}\ \emph
  {et~al.}(2016{\natexlab{a}})\citenamefont {Flores-Livas}, \citenamefont
  {Sanna},\ and\ \citenamefont {Gross}}]{Flores-Livas_H3Se2016}%
  \BibitemOpen
  \bibfield  {author} {\bibinfo {author} {\bibfnamefont {A.~J.}\ \bibnamefont
  {Flores-Livas}}, \bibinfo {author} {\bibfnamefont {A.}~\bibnamefont {Sanna}},
  \ and\ \bibinfo {author} {\bibfnamefont {E.}~\bibnamefont {Gross}},\ }\href
  {\doibase 10.1140/epjb/e2016-70020-0} {\bibfield  {journal} {\bibinfo
  {journal} {Eur. Phys. J. B}\ }\textbf {\bibinfo {volume} {89}},\ \bibinfo
  {pages} {1} (\bibinfo {year} {2016}{\natexlab{a}})}\BibitemShut {NoStop}%
\bibitem [{\citenamefont {Ashcroft}(1968)}]{Ashcroft_PRL1968}%
  \BibitemOpen
  \bibfield  {author} {\bibinfo {author} {\bibfnamefont {N.}~\bibnamefont
  {Ashcroft}},\ }\href {\doibase 10.1103/PhysRevLett.21.1748} {\bibfield
  {journal} {\bibinfo  {journal} {Phys. Rev. Lett.}\ }\textbf {\bibinfo
  {volume} {21}},\ \bibinfo {pages} {1748} (\bibinfo {year}
  {1968})}\BibitemShut {NoStop}%
\bibitem [{\citenamefont {Richardson}\ and\ \citenamefont
  {Ashcroft}(1997)}]{RichardsonAshcroft_PRL97}%
  \BibitemOpen
  \bibfield  {author} {\bibinfo {author} {\bibfnamefont {C.~F.}\ \bibnamefont
  {Richardson}}\ and\ \bibinfo {author} {\bibfnamefont {N.~W.}\ \bibnamefont
  {Ashcroft}},\ }\href {\doibase 10.1103/PhysRevLett.78.118} {\bibfield
  {journal} {\bibinfo  {journal} {Phys. Rev. Lett.}\ }\textbf {\bibinfo
  {volume} {78}},\ \bibinfo {pages} {118} (\bibinfo {year} {1997})}\BibitemShut
  {NoStop}%
\bibitem [{\citenamefont {Cudazzo}\ \emph {et~al.}(2008)\citenamefont
  {Cudazzo}, \citenamefont {Profeta}, \citenamefont {Sanna}, \citenamefont
  {Floris}, \citenamefont {Continenza}, \citenamefont {Massidda},\ and\
  \citenamefont {Gross}}]{Cudazzo_PRL2008}%
  \BibitemOpen
  \bibfield  {author} {\bibinfo {author} {\bibfnamefont {P.}~\bibnamefont
  {Cudazzo}}, \bibinfo {author} {\bibfnamefont {G.}~\bibnamefont {Profeta}},
  \bibinfo {author} {\bibfnamefont {A.}~\bibnamefont {Sanna}}, \bibinfo
  {author} {\bibfnamefont {A.}~\bibnamefont {Floris}}, \bibinfo {author}
  {\bibfnamefont {A.}~\bibnamefont {Continenza}}, \bibinfo {author}
  {\bibfnamefont {S.}~\bibnamefont {Massidda}}, \ and\ \bibinfo {author}
  {\bibfnamefont {E.~K.~U.}\ \bibnamefont {Gross}},\ }\href {\doibase
  10.1103/PhysRevLett.100.257001} {\bibfield  {journal} {\bibinfo  {journal}
  {Phys. Rev. Lett.}\ }\textbf {\bibinfo {volume} {100}},\ \bibinfo {pages}
  {257001} (\bibinfo {year} {2008})}\BibitemShut {NoStop}%
\bibitem [{\citenamefont {McMahon}\ and\ \citenamefont
  {Ceperley}(2011)}]{mcmahon_high_2011}%
  \BibitemOpen
  \bibfield  {author} {\bibinfo {author} {\bibfnamefont {J.~M.}\ \bibnamefont
  {McMahon}}\ and\ \bibinfo {author} {\bibfnamefont {D.~M.}\ \bibnamefont
  {Ceperley}},\ }\href {\doibase 10.1103/PhysRevB.84.144515} {\bibfield
  {journal} {\bibinfo  {journal} {Phys. Rev. B}\ }\textbf {\bibinfo {volume}
  {84}},\ \bibinfo {pages} {144515} (\bibinfo {year} {2011})}\BibitemShut
  {NoStop}%
\bibitem [{\citenamefont {McMahon}\ and\ \citenamefont
  {Ceperley}(2012)}]{PRB_H-tc_erratum}%
  \BibitemOpen
  \bibfield  {author} {\bibinfo {author} {\bibfnamefont {J.~M.}\ \bibnamefont
  {McMahon}}\ and\ \bibinfo {author} {\bibfnamefont {D.~M.}\ \bibnamefont
  {Ceperley}},\ }\href {\doibase 10.1103/PhysRevB.85.219902} {\bibfield
  {journal} {\bibinfo  {journal} {Phys. Rev. B}\ }\textbf {\bibinfo {volume}
  {85}},\ \bibinfo {pages} {219902(E)} (\bibinfo {year} {2012})}\BibitemShut
  {NoStop}%
\bibitem [{\citenamefont {Ashcroft}(2004)}]{Ashcroft_PRL2004}%
  \BibitemOpen
  \bibfield  {author} {\bibinfo {author} {\bibfnamefont {N.~W.}\ \bibnamefont
  {Ashcroft}},\ }\href {\doibase 10.1103/PhysRevLett.92.187002} {\bibfield
  {journal} {\bibinfo  {journal} {Phys. Rev. Lett.}\ }\textbf {\bibinfo
  {volume} {92}},\ \bibinfo {pages} {187002} (\bibinfo {year}
  {2004})}\BibitemShut {NoStop}%
\bibitem [{\citenamefont {Tse}\ \emph {et~al.}(2007)\citenamefont {Tse},
  \citenamefont {Yao},\ and\ \citenamefont {Tanaka}}]{tse_novel_2007}%
  \BibitemOpen
  \bibfield  {author} {\bibinfo {author} {\bibfnamefont {J.~S.}\ \bibnamefont
  {Tse}}, \bibinfo {author} {\bibfnamefont {Y.}~\bibnamefont {Yao}}, \ and\
  \bibinfo {author} {\bibfnamefont {K.}~\bibnamefont {Tanaka}},\ }\href
  {\doibase 10.1103/PhysRevLett.98.117004} {\bibfield  {journal} {\bibinfo
  {journal} {Phy. Rev. Lett.}\ }\textbf {\bibinfo {volume} {98}},\ \bibinfo
  {pages} {117004} (\bibinfo {year} {2007})}\BibitemShut {NoStop}%
\bibitem [{\citenamefont {Chen}\ \emph {et~al.}(2008)\citenamefont {Chen},
  \citenamefont {Struzhkin}, \citenamefont {Song}, \citenamefont {Goncharov},
  \citenamefont {Ahart}, \citenamefont {Liu}, \citenamefont {Mao},\ and\
  \citenamefont {Hemley}}]{Chen_PNAS2008}%
  \BibitemOpen
  \bibfield  {author} {\bibinfo {author} {\bibfnamefont {X.-J.}\ \bibnamefont
  {Chen}}, \bibinfo {author} {\bibfnamefont {V.~V.}\ \bibnamefont {Struzhkin}},
  \bibinfo {author} {\bibfnamefont {Y.}~\bibnamefont {Song}}, \bibinfo {author}
  {\bibfnamefont {A.~F.}\ \bibnamefont {Goncharov}}, \bibinfo {author}
  {\bibfnamefont {M.}~\bibnamefont {Ahart}}, \bibinfo {author} {\bibfnamefont
  {Z.}~\bibnamefont {Liu}}, \bibinfo {author} {\bibfnamefont {H.-k.}\
  \bibnamefont {Mao}}, \ and\ \bibinfo {author} {\bibfnamefont {R.~J.}\
  \bibnamefont {Hemley}},\ }\href {\doibase 10.1073/pnas.0710473105} {\bibfield
   {journal} {\bibinfo  {journal} {Proc. Natl. Acad. Sci. U. S. A.}\ }\textbf
  {\bibinfo {volume} {105}},\ \bibinfo {pages} {20} (\bibinfo {year}
  {2008})}\BibitemShut {NoStop}%
\bibitem [{\citenamefont {Kim}\ \emph {et~al.}(2008)\citenamefont {Kim},
  \citenamefont {Scheicher}, \citenamefont {Leb\`egue}, \citenamefont
  {Prasongkit}, \citenamefont {Arnaud}, \citenamefont {Alouani},\ and\
  \citenamefont {Ahuja}}]{Kim_PNAS2008}%
  \BibitemOpen
  \bibfield  {author} {\bibinfo {author} {\bibfnamefont {D.~Y.}\ \bibnamefont
  {Kim}}, \bibinfo {author} {\bibfnamefont {R.~H.}\ \bibnamefont {Scheicher}},
  \bibinfo {author} {\bibfnamefont {S.}~\bibnamefont {Leb\`egue}}, \bibinfo
  {author} {\bibfnamefont {J.}~\bibnamefont {Prasongkit}}, \bibinfo {author}
  {\bibfnamefont {B.}~\bibnamefont {Arnaud}}, \bibinfo {author} {\bibfnamefont
  {M.}~\bibnamefont {Alouani}}, \ and\ \bibinfo {author} {\bibfnamefont
  {R.}~\bibnamefont {Ahuja}},\ }\href {\doibase 10.1073/pnas.0804148105}
  {\bibfield  {journal} {\bibinfo  {journal} {Proc. Natl. Acad. Sci. U. S. A.}\
  }\textbf {\bibinfo {volume} {105}},\ \bibinfo {pages} {16454} (\bibinfo
  {year} {2008})}\BibitemShut {NoStop}%
\bibitem [{\citenamefont {Feng}\ \emph {et~al.}(2008)\citenamefont {Feng},
  \citenamefont {Hennig}, \citenamefont {Ashcroft},\ and\ \citenamefont
  {Hoffmann}}]{FengAsHoffman_Nature2008}%
  \BibitemOpen
  \bibfield  {author} {\bibinfo {author} {\bibfnamefont {J.}~\bibnamefont
  {Feng}}, \bibinfo {author} {\bibfnamefont {R.~G.}\ \bibnamefont {Hennig}},
  \bibinfo {author} {\bibfnamefont {N.~W.}\ \bibnamefont {Ashcroft}}, \ and\
  \bibinfo {author} {\bibfnamefont {R.}~\bibnamefont {Hoffmann}},\ }\href
  {\doibase 10.1038/nature06442} {\bibfield  {journal} {\bibinfo  {journal}
  {Nature}\ }\textbf {\bibinfo {volume} {451}},\ \bibinfo {pages} {445}
  (\bibinfo {year} {2008})}\BibitemShut {NoStop}%
\bibitem [{\citenamefont {Wang}\ \emph {et~al.}(2009)\citenamefont {Wang},
  \citenamefont {Mao}, \citenamefont {Chen},\ and\ \citenamefont
  {Mao}}]{Wang_PNAS2009}%
  \BibitemOpen
  \bibfield  {author} {\bibinfo {author} {\bibfnamefont {S.}~\bibnamefont
  {Wang}}, \bibinfo {author} {\bibfnamefont {H.-k.}\ \bibnamefont {Mao}},
  \bibinfo {author} {\bibfnamefont {X.-J.}\ \bibnamefont {Chen}}, \ and\
  \bibinfo {author} {\bibfnamefont {W.~L.}\ \bibnamefont {Mao}},\ }\href
  {\doibase 10.1073/pnas.0907729106} {\bibfield  {journal} {\bibinfo  {journal}
  {Proc. Natl. Acad. Sci. U. S. A.}\ }\textbf {\bibinfo {volume} {106}},\
  \bibinfo {pages} {14763} (\bibinfo {year} {2009})}\BibitemShut {NoStop}%
\bibitem [{\citenamefont {Yao}\ and\ \citenamefont
  {Klug}(2010)}]{Yao_PNAS2010}%
  \BibitemOpen
  \bibfield  {author} {\bibinfo {author} {\bibfnamefont {Y.}~\bibnamefont
  {Yao}}\ and\ \bibinfo {author} {\bibfnamefont {D.~D.}\ \bibnamefont {Klug}},\
  }\href {\doibase 10.1073/pnas.1006508107} {\bibfield  {journal} {\bibinfo
  {journal} {Proc. Natl. Acad. Sci. U. S. A.}\ }\textbf {\bibinfo {volume}
  {107}},\ \bibinfo {pages} {20893} (\bibinfo {year} {2010})}\BibitemShut
  {NoStop}%
\bibitem [{\citenamefont {Gao}\ \emph {et~al.}(2010)\citenamefont {Gao},
  \citenamefont {Oganov}, \citenamefont {Li}, \citenamefont {Li}, \citenamefont
  {Wang}, \citenamefont {Cui}, \citenamefont {Ma}, \citenamefont {Bergara},
  \citenamefont {Lyakhov}, \citenamefont {Iitaka},\ and\ \citenamefont
  {Zou}}]{gao_high-pressure_2010}%
  \BibitemOpen
  \bibfield  {author} {\bibinfo {author} {\bibfnamefont {G.}~\bibnamefont
  {Gao}}, \bibinfo {author} {\bibfnamefont {A.~R.}\ \bibnamefont {Oganov}},
  \bibinfo {author} {\bibfnamefont {P.}~\bibnamefont {Li}}, \bibinfo {author}
  {\bibfnamefont {Z.}~\bibnamefont {Li}}, \bibinfo {author} {\bibfnamefont
  {H.}~\bibnamefont {Wang}}, \bibinfo {author} {\bibfnamefont {T.}~\bibnamefont
  {Cui}}, \bibinfo {author} {\bibfnamefont {Y.}~\bibnamefont {Ma}}, \bibinfo
  {author} {\bibfnamefont {A.}~\bibnamefont {Bergara}}, \bibinfo {author}
  {\bibfnamefont {A.~O.}\ \bibnamefont {Lyakhov}}, \bibinfo {author}
  {\bibfnamefont {T.}~\bibnamefont {Iitaka}}, \ and\ \bibinfo {author}
  {\bibfnamefont {G.}~\bibnamefont {Zou}},\ }\href {\doibase
  10.1073/pnas.0908342107} {\bibfield  {journal} {\bibinfo  {journal} {Proc.
  Natl. Acad. Sci. U. S. A.}\ }\textbf {\bibinfo {volume} {107}},\ \bibinfo
  {pages} {1317} (\bibinfo {year} {2010})}\BibitemShut {NoStop}%
\bibitem [{\citenamefont {Kim}\ \emph {et~al.}(2010)\citenamefont {Kim},
  \citenamefont {Scheicher}, \citenamefont {Mao}, \citenamefont {Kang},\ and\
  \citenamefont {Ahuja}}]{Kim_PNAS2010}%
  \BibitemOpen
  \bibfield  {author} {\bibinfo {author} {\bibfnamefont {D.~Y.}\ \bibnamefont
  {Kim}}, \bibinfo {author} {\bibfnamefont {R.~H.}\ \bibnamefont {Scheicher}},
  \bibinfo {author} {\bibfnamefont {H.-k.}\ \bibnamefont {Mao}}, \bibinfo
  {author} {\bibfnamefont {T.~W.}\ \bibnamefont {Kang}}, \ and\ \bibinfo
  {author} {\bibfnamefont {R.}~\bibnamefont {Ahuja}},\ }\href {\doibase
  10.1073/pnas.0914462107} {\bibfield  {journal} {\bibinfo  {journal} {Proc.
  Natl. Acad. Sci. U. S. A.}\ }\textbf {\bibinfo {volume} {107}},\ \bibinfo
  {pages} {2793} (\bibinfo {year} {2010})}\BibitemShut {NoStop}%
\bibitem [{\citenamefont {Li}\ \emph {et~al.}(2010)\citenamefont {Li},
  \citenamefont {Gao}, \citenamefont {Xie}, \citenamefont {Ma}, \citenamefont
  {Cui},\ and\ \citenamefont {Zou}}]{Li_PNAS2010}%
  \BibitemOpen
  \bibfield  {author} {\bibinfo {author} {\bibfnamefont {Y.}~\bibnamefont
  {Li}}, \bibinfo {author} {\bibfnamefont {G.}~\bibnamefont {Gao}}, \bibinfo
  {author} {\bibfnamefont {Y.}~\bibnamefont {Xie}}, \bibinfo {author}
  {\bibfnamefont {Y.}~\bibnamefont {Ma}}, \bibinfo {author} {\bibfnamefont
  {T.}~\bibnamefont {Cui}}, \ and\ \bibinfo {author} {\bibfnamefont
  {G.}~\bibnamefont {Zou}},\ }\href {\doibase 10.1073/pnas.1007354107}
  {\bibfield  {journal} {\bibinfo  {journal} {Proc. Natl. Acad. Sci. U. S. A.}\
  }\textbf {\bibinfo {volume} {107}},\ \bibinfo {pages} {15708} (\bibinfo
  {year} {2010})}\BibitemShut {NoStop}%
\bibitem [{\citenamefont {Zhou}\ \emph {et~al.}(2012)\citenamefont {Zhou},
  \citenamefont {Jin}, \citenamefont {Meng}, \citenamefont {Bao}, \citenamefont
  {Ma}, \citenamefont {Liu},\ and\ \citenamefont {Cui}}]{Zhou_PRB2012}%
  \BibitemOpen
  \bibfield  {author} {\bibinfo {author} {\bibfnamefont {D.}~\bibnamefont
  {Zhou}}, \bibinfo {author} {\bibfnamefont {X.}~\bibnamefont {Jin}}, \bibinfo
  {author} {\bibfnamefont {X.}~\bibnamefont {Meng}}, \bibinfo {author}
  {\bibfnamefont {G.}~\bibnamefont {Bao}}, \bibinfo {author} {\bibfnamefont
  {Y.}~\bibnamefont {Ma}}, \bibinfo {author} {\bibfnamefont {B.}~\bibnamefont
  {Liu}}, \ and\ \bibinfo {author} {\bibfnamefont {T.}~\bibnamefont {Cui}},\
  }\href {\doibase 10.1103/PhysRevB.86.014118} {\bibfield  {journal} {\bibinfo
  {journal} {Phys. Rev. B}\ }\textbf {\bibinfo {volume} {86}},\ \bibinfo
  {pages} {014118} (\bibinfo {year} {2012})}\BibitemShut {NoStop}%
\bibitem [{\citenamefont {Hooper}\ \emph {et~al.}(2014)\citenamefont {Hooper},
  \citenamefont {Terpstra}, \citenamefont {Shamp},\ and\ \citenamefont
  {Zurek}}]{Hooper_JPC-2014}%
  \BibitemOpen
  \bibfield  {author} {\bibinfo {author} {\bibfnamefont {J.}~\bibnamefont
  {Hooper}}, \bibinfo {author} {\bibfnamefont {T.}~\bibnamefont {Terpstra}},
  \bibinfo {author} {\bibfnamefont {A.}~\bibnamefont {Shamp}}, \ and\ \bibinfo
  {author} {\bibfnamefont {E.}~\bibnamefont {Zurek}},\ }\href {\doibase
  10.1021/jp4125342} {\bibfield  {journal} {\bibinfo  {journal} {The Journal of
  Physical Chemistry C}\ }\textbf {\bibinfo {volume} {118}},\ \bibinfo {pages}
  {6433} (\bibinfo {year} {2014})}\BibitemShut {NoStop}%
\bibitem [{\citenamefont {Esfahani}\ \emph {et~al.}(2016)\citenamefont
  {Esfahani}, \citenamefont {Wang}, \citenamefont {Oganov}, \citenamefont
  {Dong}, \citenamefont {Zhu}, \citenamefont {Wang}, \citenamefont {Rakitin},\
  and\ \citenamefont {Zhou}}]{esfahani2016superconductivity}%
  \BibitemOpen
  \bibfield  {author} {\bibinfo {author} {\bibfnamefont {M.~M.~D.}\
  \bibnamefont {Esfahani}}, \bibinfo {author} {\bibfnamefont {Z.}~\bibnamefont
  {Wang}}, \bibinfo {author} {\bibfnamefont {A.~R.}\ \bibnamefont {Oganov}},
  \bibinfo {author} {\bibfnamefont {H.}~\bibnamefont {Dong}}, \bibinfo {author}
  {\bibfnamefont {Q.}~\bibnamefont {Zhu}}, \bibinfo {author} {\bibfnamefont
  {S.}~\bibnamefont {Wang}}, \bibinfo {author} {\bibfnamefont {M.~S.}\
  \bibnamefont {Rakitin}}, \ and\ \bibinfo {author} {\bibfnamefont {X.-F.}\
  \bibnamefont {Zhou}},\ }\href@noop {} {\bibfield  {journal} {\bibinfo
  {journal} {Scientific reports}\ }\textbf {\bibinfo {volume} {6}} (\bibinfo
  {year} {2016})}\BibitemShut {NoStop}%
\bibitem [{\citenamefont {Struzhkin}\ \emph {et~al.}(2014)\citenamefont
  {Struzhkin}, \citenamefont {Kim}, \citenamefont {Stavrou}, \citenamefont
  {Muramatsu}, \citenamefont {Mao}, \citenamefont {Pickard}, \citenamefont
  {Needs}, \citenamefont {Prakapenka},\ and\ \citenamefont
  {Goncharov}}]{struzhkin2014synthesis}%
  \BibitemOpen
  \bibfield  {author} {\bibinfo {author} {\bibfnamefont {V.~V.}\ \bibnamefont
  {Struzhkin}}, \bibinfo {author} {\bibfnamefont {D.}~\bibnamefont {Kim}},
  \bibinfo {author} {\bibfnamefont {E.}~\bibnamefont {Stavrou}}, \bibinfo
  {author} {\bibfnamefont {T.}~\bibnamefont {Muramatsu}}, \bibinfo {author}
  {\bibfnamefont {H.}~\bibnamefont {Mao}}, \bibinfo {author} {\bibfnamefont
  {C.~J.}\ \bibnamefont {Pickard}}, \bibinfo {author} {\bibfnamefont {R.~J.}\
  \bibnamefont {Needs}}, \bibinfo {author} {\bibfnamefont {V.~B.}\ \bibnamefont
  {Prakapenka}}, \ and\ \bibinfo {author} {\bibfnamefont {A.~F.}\ \bibnamefont
  {Goncharov}},\ }\href@noop {} {\bibfield  {journal} {\bibinfo  {journal}
  {arXiv preprint arXiv:1412.1542}\ } (\bibinfo {year} {2014})}\BibitemShut
  {NoStop}%
\bibitem [{\citenamefont {Zhong}\ \emph {et~al.}(2016)\citenamefont {Zhong},
  \citenamefont {Wang}, \citenamefont {Zhang}, \citenamefont {Liu},
  \citenamefont {Zhang}, \citenamefont {Song}, \citenamefont {Yang},
  \citenamefont {Zhang},\ and\ \citenamefont {Ma}}]{PRL_TeH3_2016}%
  \BibitemOpen
  \bibfield  {author} {\bibinfo {author} {\bibfnamefont {X.}~\bibnamefont
  {Zhong}}, \bibinfo {author} {\bibfnamefont {H.}~\bibnamefont {Wang}},
  \bibinfo {author} {\bibfnamefont {J.}~\bibnamefont {Zhang}}, \bibinfo
  {author} {\bibfnamefont {H.}~\bibnamefont {Liu}}, \bibinfo {author}
  {\bibfnamefont {S.}~\bibnamefont {Zhang}}, \bibinfo {author} {\bibfnamefont
  {H.-F.}\ \bibnamefont {Song}}, \bibinfo {author} {\bibfnamefont
  {G.}~\bibnamefont {Yang}}, \bibinfo {author} {\bibfnamefont {L.}~\bibnamefont
  {Zhang}}, \ and\ \bibinfo {author} {\bibfnamefont {Y.}~\bibnamefont {Ma}},\
  }\href {\doibase 10.1103/PhysRevLett.116.057002} {\bibfield  {journal}
  {\bibinfo  {journal} {Phys. Rev. Lett.}\ }\textbf {\bibinfo {volume} {116}},\
  \bibinfo {pages} {057002} (\bibinfo {year} {2016})}\BibitemShut {NoStop}%
\bibitem [{\citenamefont {Eremets}\ \emph {et~al.}(2008)\citenamefont
  {Eremets}, \citenamefont {Trojan}, \citenamefont {Medvedev}, \citenamefont
  {Tse},\ and\ \citenamefont {Yao}}]{Eremets_Science2008}%
  \BibitemOpen
  \bibfield  {author} {\bibinfo {author} {\bibfnamefont {M.~I.}\ \bibnamefont
  {Eremets}}, \bibinfo {author} {\bibfnamefont {I.~A.}\ \bibnamefont {Trojan}},
  \bibinfo {author} {\bibfnamefont {S.~A.}\ \bibnamefont {Medvedev}}, \bibinfo
  {author} {\bibfnamefont {J.~S.}\ \bibnamefont {Tse}}, \ and\ \bibinfo
  {author} {\bibfnamefont {Y.}~\bibnamefont {Yao}},\ }\href {\doibase
  10.1126/science.1153282} {\bibfield  {journal} {\bibinfo  {journal}
  {Science}\ }\textbf {\bibinfo {volume} {319}},\ \bibinfo {pages} {1506}
  (\bibinfo {year} {2008})}\BibitemShut {NoStop}%
\bibitem [{\citenamefont {Degtyareva}\ \emph {et~al.}(2009)\citenamefont
  {Degtyareva}, \citenamefont {Proctor}, \citenamefont {Guillaume},
  \citenamefont {Gregoryanz},\ and\ \citenamefont
  {Hanfland}}]{degtyareva_formation_2009}%
  \BibitemOpen
  \bibfield  {author} {\bibinfo {author} {\bibfnamefont {O.}~\bibnamefont
  {Degtyareva}}, \bibinfo {author} {\bibfnamefont {J.~E.}\ \bibnamefont
  {Proctor}}, \bibinfo {author} {\bibfnamefont {C.~L.}\ \bibnamefont
  {Guillaume}}, \bibinfo {author} {\bibfnamefont {E.}~\bibnamefont
  {Gregoryanz}}, \ and\ \bibinfo {author} {\bibfnamefont {M.}~\bibnamefont
  {Hanfland}},\ }\href {\doibase 10.1016/j.ssc.2009.07.022} {\bibfield
  {journal} {\bibinfo  {journal} {Solid State Commun.}\ }\textbf {\bibinfo
  {volume} {149}},\ \bibinfo {pages} {1583} (\bibinfo {year}
  {2009})}\BibitemShut {NoStop}%
\bibitem [{\citenamefont {Hanfland}\ \emph {et~al.}(2011)\citenamefont
  {Hanfland}, \citenamefont {Proctor}, \citenamefont {Guillaume}, \citenamefont
  {Degtyareva},\ and\ \citenamefont {Gregoryanz}}]{Hanfland_PRL2011}%
  \BibitemOpen
  \bibfield  {author} {\bibinfo {author} {\bibfnamefont {M.}~\bibnamefont
  {Hanfland}}, \bibinfo {author} {\bibfnamefont {J.~E.}\ \bibnamefont
  {Proctor}}, \bibinfo {author} {\bibfnamefont {C.~L.}\ \bibnamefont
  {Guillaume}}, \bibinfo {author} {\bibfnamefont {O.}~\bibnamefont
  {Degtyareva}}, \ and\ \bibinfo {author} {\bibfnamefont {E.}~\bibnamefont
  {Gregoryanz}},\ }\href {\doibase 10.1103/PhysRevLett.106.095503} {\bibfield
  {journal} {\bibinfo  {journal} {Phy. Rev. Lett.}\ }\textbf {\bibinfo {volume}
  {106}},\ \bibinfo {pages} {095503} (\bibinfo {year} {2011})}\BibitemShut
  {NoStop}%
\bibitem [{\citenamefont {Strobel}\ \emph
  {et~al.}(2011{\natexlab{a}})\citenamefont {Strobel}, \citenamefont {Ganesh},
  \citenamefont {Somayazulu}, \citenamefont {Kent},\ and\ \citenamefont
  {Hemley}}]{Strobel_PRL2011}%
  \BibitemOpen
  \bibfield  {author} {\bibinfo {author} {\bibfnamefont {T.~A.}\ \bibnamefont
  {Strobel}}, \bibinfo {author} {\bibfnamefont {P.}~\bibnamefont {Ganesh}},
  \bibinfo {author} {\bibfnamefont {M.}~\bibnamefont {Somayazulu}}, \bibinfo
  {author} {\bibfnamefont {P.~R.~C.}\ \bibnamefont {Kent}}, \ and\ \bibinfo
  {author} {\bibfnamefont {R.~J.}\ \bibnamefont {Hemley}},\ }\href {\doibase
  10.1103/PhysRevLett.107.255503} {\bibfield  {journal} {\bibinfo  {journal}
  {Phys. Rev. Lett.}\ }\textbf {\bibinfo {volume} {107}},\ \bibinfo {pages}
  {255503} (\bibinfo {year} {2011}{\natexlab{a}})}\BibitemShut {NoStop}%
\bibitem [{\citenamefont {Flores-Livas}\ \emph {et~al.}(2012)\citenamefont
  {Flores-Livas}, \citenamefont {Amsler}, \citenamefont {Lenosky},
  \citenamefont {Lehtovaara}, \citenamefont {Botti}, \citenamefont {Marques},\
  and\ \citenamefont {Goedecker}}]{Disilane_JAFL}%
  \BibitemOpen
  \bibfield  {author} {\bibinfo {author} {\bibfnamefont {J.~A.}\ \bibnamefont
  {Flores-Livas}}, \bibinfo {author} {\bibfnamefont {M.}~\bibnamefont
  {Amsler}}, \bibinfo {author} {\bibfnamefont {T.~J.}\ \bibnamefont {Lenosky}},
  \bibinfo {author} {\bibfnamefont {L.}~\bibnamefont {Lehtovaara}}, \bibinfo
  {author} {\bibfnamefont {S.}~\bibnamefont {Botti}}, \bibinfo {author}
  {\bibfnamefont {M.~A.~L.}\ \bibnamefont {Marques}}, \ and\ \bibinfo {author}
  {\bibfnamefont {S.}~\bibnamefont {Goedecker}},\ }\href {\doibase
  10.1103/PhysRevLett.108.117004} {\bibfield  {journal} {\bibinfo  {journal}
  {Phys. Rev. Lett.}\ }\textbf {\bibinfo {volume} {108}},\ \bibinfo {pages}
  {117004} (\bibinfo {year} {2012})}\BibitemShut {NoStop}%
\bibitem [{\citenamefont {Errea}\ \emph {et~al.}(2016)\citenamefont {Errea},
  \citenamefont {Calandra}, \citenamefont {Pickard}, \citenamefont {Nelson},
  \citenamefont {Needs}, \citenamefont {Li}, \citenamefont {Liu}, \citenamefont
  {Zhang}, \citenamefont {Ma},\ and\ \citenamefont {Mauri}}]{errea2016quantum}%
  \BibitemOpen
  \bibfield  {author} {\bibinfo {author} {\bibfnamefont {I.}~\bibnamefont
  {Errea}}, \bibinfo {author} {\bibfnamefont {M.}~\bibnamefont {Calandra}},
  \bibinfo {author} {\bibfnamefont {C.~J.}\ \bibnamefont {Pickard}}, \bibinfo
  {author} {\bibfnamefont {J.~R.}\ \bibnamefont {Nelson}}, \bibinfo {author}
  {\bibfnamefont {R.~J.}\ \bibnamefont {Needs}}, \bibinfo {author}
  {\bibfnamefont {Y.}~\bibnamefont {Li}}, \bibinfo {author} {\bibfnamefont
  {H.}~\bibnamefont {Liu}}, \bibinfo {author} {\bibfnamefont {Y.}~\bibnamefont
  {Zhang}}, \bibinfo {author} {\bibfnamefont {Y.}~\bibnamefont {Ma}}, \ and\
  \bibinfo {author} {\bibfnamefont {F.}~\bibnamefont {Mauri}},\ }\href@noop {}
  {\bibfield  {journal} {\bibinfo  {journal} {Nature}\ }\textbf {\bibinfo
  {volume} {532}},\ \bibinfo {pages} {81} (\bibinfo {year} {2016})}\BibitemShut
  {NoStop}%
\bibitem [{\citenamefont {Errea}\ \emph {et~al.}(2015)\citenamefont {Errea},
  \citenamefont {Calandra}, \citenamefont {Pickard}, \citenamefont {Nelson},
  \citenamefont {Needs}, \citenamefont {Li}, \citenamefont {Liu}, \citenamefont
  {Zhang}, \citenamefont {Ma},\ and\ \citenamefont
  {Mauri}}]{Errea_anhaPRL2015}%
  \BibitemOpen
  \bibfield  {author} {\bibinfo {author} {\bibfnamefont {I.}~\bibnamefont
  {Errea}}, \bibinfo {author} {\bibfnamefont {M.}~\bibnamefont {Calandra}},
  \bibinfo {author} {\bibfnamefont {C.~J.}\ \bibnamefont {Pickard}}, \bibinfo
  {author} {\bibfnamefont {J.}~\bibnamefont {Nelson}}, \bibinfo {author}
  {\bibfnamefont {R.~J.}\ \bibnamefont {Needs}}, \bibinfo {author}
  {\bibfnamefont {Y.}~\bibnamefont {Li}}, \bibinfo {author} {\bibfnamefont
  {H.}~\bibnamefont {Liu}}, \bibinfo {author} {\bibfnamefont {Y.}~\bibnamefont
  {Zhang}}, \bibinfo {author} {\bibfnamefont {Y.}~\bibnamefont {Ma}}, \ and\
  \bibinfo {author} {\bibfnamefont {F.}~\bibnamefont {Mauri}},\ }\href
  {\doibase 10.1103/PhysRevLett.114.157004} {\bibfield  {journal} {\bibinfo
  {journal} {Phys. Rev. Lett.}\ }\textbf {\bibinfo {volume} {114}},\ \bibinfo
  {pages} {157004} (\bibinfo {year} {2015})}\BibitemShut {NoStop}%
\bibitem [{\citenamefont {Akashi}\ \emph {et~al.}(2015)\citenamefont {Akashi},
  \citenamefont {Kawamura}, \citenamefont {Tsuneyuki}, \citenamefont {Nomura},\
  and\ \citenamefont {Arita}}]{PRB_akashi_2015_HS}%
  \BibitemOpen
  \bibfield  {author} {\bibinfo {author} {\bibfnamefont {R.}~\bibnamefont
  {Akashi}}, \bibinfo {author} {\bibfnamefont {M.}~\bibnamefont {Kawamura}},
  \bibinfo {author} {\bibfnamefont {S.}~\bibnamefont {Tsuneyuki}}, \bibinfo
  {author} {\bibfnamefont {Y.}~\bibnamefont {Nomura}}, \ and\ \bibinfo {author}
  {\bibfnamefont {R.}~\bibnamefont {Arita}},\ }\href {\doibase
  10.1103/PhysRevB.91.224513} {\bibfield  {journal} {\bibinfo  {journal} {Phys.
  Rev. B}\ }\textbf {\bibinfo {volume} {91}},\ \bibinfo {pages} {224513}
  (\bibinfo {year} {2015})}\BibitemShut {NoStop}%
\bibitem [{\citenamefont {Quan}\ and\ \citenamefont
  {Pickett}(2015)}]{quan_impact_2015}%
  \BibitemOpen
  \bibfield  {author} {\bibinfo {author} {\bibfnamefont {Y.}~\bibnamefont
  {Quan}}\ and\ \bibinfo {author} {\bibfnamefont {W.~E.}\ \bibnamefont
  {Pickett}},\ }\href@noop {} {\bibfield  {journal} {\bibinfo  {journal}
  {arXiv:~1508.04491 [cond-mat.supr-con]}\ } (\bibinfo {year}
  {2015})}\BibitemShut {NoStop}%
\bibitem [{\citenamefont {Xie}\ \emph {et~al.}(2014)\citenamefont {Xie},
  \citenamefont {Li}, \citenamefont {Oganov},\ and\ \citenamefont
  {Wang}}]{xie2014superconductivity}%
  \BibitemOpen
  \bibfield  {author} {\bibinfo {author} {\bibfnamefont {Y.}~\bibnamefont
  {Xie}}, \bibinfo {author} {\bibfnamefont {Q.}~\bibnamefont {Li}}, \bibinfo
  {author} {\bibfnamefont {A.~R.}\ \bibnamefont {Oganov}}, \ and\ \bibinfo
  {author} {\bibfnamefont {H.}~\bibnamefont {Wang}},\ }\href@noop {} {\bibfield
   {journal} {\bibinfo  {journal} {Acta Crystallographica Section C: Structural
  Chemistry}\ }\textbf {\bibinfo {volume} {70}},\ \bibinfo {pages} {104}
  (\bibinfo {year} {2014})}\BibitemShut {NoStop}%
\bibitem [{\citenamefont {Ortenzi}\ \emph {et~al.}(2015)\citenamefont
  {Ortenzi}, \citenamefont {Cappelluti},\ and\ \citenamefont
  {Pietronero}}]{ortenzi_TB_2015}%
  \BibitemOpen
  \bibfield  {author} {\bibinfo {author} {\bibfnamefont {L.}~\bibnamefont
  {Ortenzi}}, \bibinfo {author} {\bibfnamefont {E.}~\bibnamefont {Cappelluti}},
  \ and\ \bibinfo {author} {\bibfnamefont {L.}~\bibnamefont {Pietronero}},\
  }\href@noop {} {\bibfield  {journal} {\bibinfo  {journal} {arXiv:~1511.04304
  [cond-mat.mtrl-sci]}\ } (\bibinfo {year} {2015})}\BibitemShut {NoStop}%
\bibitem [{\citenamefont {Akashi}\ \emph {et~al.}(2016)\citenamefont {Akashi},
  \citenamefont {Sano}, \citenamefont {Arita},\ and\ \citenamefont
  {Tsuneyuki}}]{akashi_mangeli-phases}%
  \BibitemOpen
  \bibfield  {author} {\bibinfo {author} {\bibfnamefont {R.}~\bibnamefont
  {Akashi}}, \bibinfo {author} {\bibfnamefont {W.}~\bibnamefont {Sano}},
  \bibinfo {author} {\bibfnamefont {R.}~\bibnamefont {Arita}}, \ and\ \bibinfo
  {author} {\bibfnamefont {S.}~\bibnamefont {Tsuneyuki}},\ }\href {\doibase
  10.1103/PhysRevLett.117.075503} {\bibfield  {journal} {\bibinfo  {journal}
  {Phys. Rev. Lett.}\ }\textbf {\bibinfo {volume} {117}},\ \bibinfo {pages}
  {075503} (\bibinfo {year} {2016})}\BibitemShut {NoStop}%
\bibitem [{\citenamefont {Drozdov}\ \emph
  {et~al.}(2015{\natexlab{b}})\citenamefont {Drozdov}, \citenamefont
  {Eremets},\ and\ \citenamefont {Troyan}}]{Drozdov_ph3_arxiv2015}%
  \BibitemOpen
  \bibfield  {author} {\bibinfo {author} {\bibfnamefont {A.}~\bibnamefont
  {Drozdov}}, \bibinfo {author} {\bibfnamefont {M.~I.}\ \bibnamefont
  {Eremets}}, \ and\ \bibinfo {author} {\bibfnamefont {I.~A.}\ \bibnamefont
  {Troyan}},\ }\href@noop {} {\bibfield  {journal} {\bibinfo  {journal}
  {arXiv:~1508.06224 [cond-mat.supr-con]}\ } (\bibinfo {year}
  {2015}{\natexlab{b}})}\BibitemShut {NoStop}%
\bibitem [{\citenamefont {Flores-Livas}\ \emph
  {et~al.}(2016{\natexlab{b}})\citenamefont {Flores-Livas}, \citenamefont
  {Amsler}, \citenamefont {Heil}, \citenamefont {Sanna}, \citenamefont {Boeri},
  \citenamefont {Profeta}, \citenamefont {Wolverton}, \citenamefont
  {Goedecker},\ and\ \citenamefont {Gross}}]{ours_PH_rPRB2016}%
  \BibitemOpen
  \bibfield  {author} {\bibinfo {author} {\bibfnamefont {J.~A.}\ \bibnamefont
  {Flores-Livas}}, \bibinfo {author} {\bibfnamefont {M.}~\bibnamefont
  {Amsler}}, \bibinfo {author} {\bibfnamefont {C.}~\bibnamefont {Heil}},
  \bibinfo {author} {\bibfnamefont {A.}~\bibnamefont {Sanna}}, \bibinfo
  {author} {\bibfnamefont {L.}~\bibnamefont {Boeri}}, \bibinfo {author}
  {\bibfnamefont {G.}~\bibnamefont {Profeta}}, \bibinfo {author} {\bibfnamefont
  {C.}~\bibnamefont {Wolverton}}, \bibinfo {author} {\bibfnamefont
  {S.}~\bibnamefont {Goedecker}}, \ and\ \bibinfo {author} {\bibfnamefont
  {E.~K.~U.}\ \bibnamefont {Gross}},\ }\href {\doibase
  10.1103/PhysRevB.93.020508} {\bibfield  {journal} {\bibinfo  {journal} {Phys.
  Rev. B}\ }\textbf {\bibinfo {volume} {93}},\ \bibinfo {pages} {020508}
  (\bibinfo {year} {2016}{\natexlab{b}})}\BibitemShut {NoStop}%
\bibitem [{\citenamefont {Wigner}\ and\ \citenamefont
  {Huntington}(1935)}]{Wigner_JCP1935}%
  \BibitemOpen
  \bibfield  {author} {\bibinfo {author} {\bibfnamefont {E.}~\bibnamefont
  {Wigner}}\ and\ \bibinfo {author} {\bibfnamefont {H.~B.}\ \bibnamefont
  {Huntington}},\ }\href {\doibase http://dx.doi.org/10.1063/1.1749590}
  {\bibfield  {journal} {\bibinfo  {journal} {J. Chem. Phys.}\ }\textbf
  {\bibinfo {volume} {3}},\ \bibinfo {pages} {764} (\bibinfo {year}
  {1935})}\BibitemShut {NoStop}%
\bibitem [{\citenamefont {Pickard}\ and\ \citenamefont
  {Needs}(2007)}]{pickard_structure_2007}%
  \BibitemOpen
  \bibfield  {author} {\bibinfo {author} {\bibfnamefont {C.~J.}\ \bibnamefont
  {Pickard}}\ and\ \bibinfo {author} {\bibfnamefont {R.~J.}\ \bibnamefont
  {Needs}},\ }\href {\doibase 10.1038/nphys625} {\bibfield  {journal} {\bibinfo
   {journal} {Nat. Phys.}\ }\textbf {\bibinfo {volume} {3}},\ \bibinfo {pages}
  {473} (\bibinfo {year} {2007})}\BibitemShut {NoStop}%
\bibitem [{\citenamefont {Loubeyre}\ \emph {et~al.}(2002)\citenamefont
  {Loubeyre}, \citenamefont {Occelli},\ and\ \citenamefont
  {LeToullec}}]{LeToullec2002}%
  \BibitemOpen
  \bibfield  {author} {\bibinfo {author} {\bibfnamefont {P.}~\bibnamefont
  {Loubeyre}}, \bibinfo {author} {\bibfnamefont {F.}~\bibnamefont {Occelli}}, \
  and\ \bibinfo {author} {\bibfnamefont {R.}~\bibnamefont {LeToullec}},\ }\href
  {\doibase 10.1038/416613a} {\bibfield  {journal} {\bibinfo  {journal}
  {Nature}\ }\textbf {\bibinfo {volume} {416}},\ \bibinfo {pages} {13}
  (\bibinfo {year} {2002})}\BibitemShut {NoStop}%
\bibitem [{\citenamefont {Eremets}\ and\ \citenamefont
  {Troyan}(2011)}]{Eremets_NatMat2011}%
  \BibitemOpen
  \bibfield  {author} {\bibinfo {author} {\bibfnamefont {M.~I.}\ \bibnamefont
  {Eremets}}\ and\ \bibinfo {author} {\bibfnamefont {I.~A.}\ \bibnamefont
  {Troyan}},\ }\href {\doibase http://dx.doi.org/10.1038/nmat3175} {\bibfield
  {journal} {\bibinfo  {journal} {Nat. Mat.}\ }\textbf {\bibinfo {volume}
  {10}},\ \bibinfo {pages} {927} (\bibinfo {year} {2011})}\BibitemShut
  {NoStop}%
\bibitem [{\citenamefont {Zha}\ \emph {et~al.}(2012)\citenamefont {Zha},
  \citenamefont {Liu},\ and\ \citenamefont {Hemley}}]{Hemley_PRL2012}%
  \BibitemOpen
  \bibfield  {author} {\bibinfo {author} {\bibfnamefont {C.-S.}\ \bibnamefont
  {Zha}}, \bibinfo {author} {\bibfnamefont {Z.}~\bibnamefont {Liu}}, \ and\
  \bibinfo {author} {\bibfnamefont {R.~J.}\ \bibnamefont {Hemley}},\ }\href
  {\doibase 10.1103/PhysRevLett.108.146402} {\bibfield  {journal} {\bibinfo
  {journal} {Phys. Rev. Lett.}\ }\textbf {\bibinfo {volume} {108}},\ \bibinfo
  {pages} {146402} (\bibinfo {year} {2012})}\BibitemShut {NoStop}%
\bibitem [{\citenamefont {Naumov}\ and\ \citenamefont
  {Hemley}(2014)}]{HRussell_hydrogenJACS2014}%
  \BibitemOpen
  \bibfield  {author} {\bibinfo {author} {\bibfnamefont {I.~I.}\ \bibnamefont
  {Naumov}}\ and\ \bibinfo {author} {\bibfnamefont {R.~J.}\ \bibnamefont
  {Hemley}},\ }\href {\doibase 10.1021/ar5002654} {\bibfield  {journal}
  {\bibinfo  {journal} {Acc. Chem. Res.}\ }\textbf {\bibinfo {volume} {47}},\
  \bibinfo {pages} {3551} (\bibinfo {year} {2014})}\BibitemShut {NoStop}%
\bibitem [{\citenamefont {Ekimov}\ \emph {et~al.}(2004)\citenamefont {Ekimov},
  \citenamefont {Sidorov}, \citenamefont {Bauer}, \citenamefont {Mel'nik},
  \citenamefont {Curro}, \citenamefont {Thompson},\ and\ \citenamefont
  {Stishov}}]{ekimov_superconductivity_2004}%
  \BibitemOpen
  \bibfield  {author} {\bibinfo {author} {\bibfnamefont {E.~A.}\ \bibnamefont
  {Ekimov}}, \bibinfo {author} {\bibfnamefont {V.~A.}\ \bibnamefont {Sidorov}},
  \bibinfo {author} {\bibfnamefont {E.~D.}\ \bibnamefont {Bauer}}, \bibinfo
  {author} {\bibfnamefont {N.~N.}\ \bibnamefont {Mel'nik}}, \bibinfo {author}
  {\bibfnamefont {N.~J.}\ \bibnamefont {Curro}}, \bibinfo {author}
  {\bibfnamefont {J.~D.}\ \bibnamefont {Thompson}}, \ and\ \bibinfo {author}
  {\bibfnamefont {S.~M.}\ \bibnamefont {Stishov}},\ }\href {\doibase
  10.1038/nature02449} {\bibfield  {journal} {\bibinfo  {journal} {Nature}\
  }\textbf {\bibinfo {volume} {428}},\ \bibinfo {pages} {542} (\bibinfo {year}
  {2004})}\BibitemShut {NoStop}%
\bibitem [{\citenamefont {Bustarret}\ \emph {et~al.}(2006)\citenamefont
  {Bustarret}, \citenamefont {Marcenat}, \citenamefont {Achatz}, \citenamefont
  {Ka{\v{c}}mar{\v{c}}ik}, \citenamefont {L{\'e}vy}, \citenamefont {Huxley},
  \citenamefont {Ort{\'e}ga}, \citenamefont {Bourgeois}, \citenamefont {Blase},
  \citenamefont {D{\'e}barre} \emph
  {et~al.}}]{bustarret_superconductivity_2006}%
  \BibitemOpen
  \bibfield  {author} {\bibinfo {author} {\bibfnamefont {E.}~\bibnamefont
  {Bustarret}}, \bibinfo {author} {\bibfnamefont {C.}~\bibnamefont {Marcenat}},
  \bibinfo {author} {\bibfnamefont {P.}~\bibnamefont {Achatz}}, \bibinfo
  {author} {\bibfnamefont {J.}~\bibnamefont {Ka{\v{c}}mar{\v{c}}ik}}, \bibinfo
  {author} {\bibfnamefont {F.}~\bibnamefont {L{\'e}vy}}, \bibinfo {author}
  {\bibfnamefont {A.}~\bibnamefont {Huxley}}, \bibinfo {author} {\bibfnamefont
  {L.}~\bibnamefont {Ort{\'e}ga}}, \bibinfo {author} {\bibfnamefont
  {E.}~\bibnamefont {Bourgeois}}, \bibinfo {author} {\bibfnamefont
  {X.}~\bibnamefont {Blase}}, \bibinfo {author} {\bibfnamefont
  {D.}~\bibnamefont {D{\'e}barre}},  \emph {et~al.},\ }\href@noop {} {\bibfield
   {journal} {\bibinfo  {journal} {Nature}\ }\textbf {\bibinfo {volume}
  {444}},\ \bibinfo {pages} {465} (\bibinfo {year} {2006})}\BibitemShut
  {NoStop}%
\bibitem [{\citenamefont {Herrmannsd\"orfer}\ \emph {et~al.}(2009)\citenamefont
  {Herrmannsd\"orfer}, \citenamefont {Heera}, \citenamefont {Ignatchik},
  \citenamefont {Uhlarz}, \citenamefont {M\"ucklich}, \citenamefont {Posselt},
  \citenamefont {Reuther}, \citenamefont {Schmidt}, \citenamefont {Heinig},
  \citenamefont {Skorupa}, \citenamefont {Voelskow}, \citenamefont
  {W\"undisch}, \citenamefont {Skrotzki}, \citenamefont {Helm},\ and\
  \citenamefont {Wosnitza}}]{herrmannsdorfer_superconducting_2009}%
  \BibitemOpen
  \bibfield  {author} {\bibinfo {author} {\bibfnamefont {T.}~\bibnamefont
  {Herrmannsd\"orfer}}, \bibinfo {author} {\bibfnamefont {V.}~\bibnamefont
  {Heera}}, \bibinfo {author} {\bibfnamefont {O.}~\bibnamefont {Ignatchik}},
  \bibinfo {author} {\bibfnamefont {M.}~\bibnamefont {Uhlarz}}, \bibinfo
  {author} {\bibfnamefont {A.}~\bibnamefont {M\"ucklich}}, \bibinfo {author}
  {\bibfnamefont {M.}~\bibnamefont {Posselt}}, \bibinfo {author} {\bibfnamefont
  {H.}~\bibnamefont {Reuther}}, \bibinfo {author} {\bibfnamefont
  {B.}~\bibnamefont {Schmidt}}, \bibinfo {author} {\bibfnamefont {K.-H.}\
  \bibnamefont {Heinig}}, \bibinfo {author} {\bibfnamefont {W.}~\bibnamefont
  {Skorupa}}, \bibinfo {author} {\bibfnamefont {M.}~\bibnamefont {Voelskow}},
  \bibinfo {author} {\bibfnamefont {C.}~\bibnamefont {W\"undisch}}, \bibinfo
  {author} {\bibfnamefont {R.}~\bibnamefont {Skrotzki}}, \bibinfo {author}
  {\bibfnamefont {M.}~\bibnamefont {Helm}}, \ and\ \bibinfo {author}
  {\bibfnamefont {J.}~\bibnamefont {Wosnitza}},\ }\href {\doibase
  10.1103/PhysRevLett.102.217003} {\bibfield  {journal} {\bibinfo  {journal}
  {Phys. Rev. Lett.}\ }\textbf {\bibinfo {volume} {102}},\ \bibinfo {pages}
  {217003} (\bibinfo {year} {2009})}\BibitemShut {NoStop}%
\bibitem [{\citenamefont {Kriener}\ \emph {et~al.}(2008)\citenamefont
  {Kriener}, \citenamefont {Muranaka}, \citenamefont {Kato}, \citenamefont
  {Ren}, \citenamefont {Akimitsu},\ and\ \citenamefont
  {Maeno}}]{kriener_superconductivity_2008}%
  \BibitemOpen
  \bibfield  {author} {\bibinfo {author} {\bibfnamefont {M.}~\bibnamefont
  {Kriener}}, \bibinfo {author} {\bibfnamefont {T.}~\bibnamefont {Muranaka}},
  \bibinfo {author} {\bibfnamefont {J.}~\bibnamefont {Kato}}, \bibinfo {author}
  {\bibfnamefont {Z.-A.}\ \bibnamefont {Ren}}, \bibinfo {author} {\bibfnamefont
  {J.}~\bibnamefont {Akimitsu}}, \ and\ \bibinfo {author} {\bibfnamefont
  {Y.}~\bibnamefont {Maeno}},\ }\href {\doibase 10.1088/1468-6996/9/4/044205}
  {\bibfield  {journal} {\bibinfo  {journal} {Sci. Technol. Adv. Mater.}\
  }\textbf {\bibinfo {volume} {9}},\ \bibinfo {pages} {044205} (\bibinfo {year}
  {2008})}\BibitemShut {NoStop}%
\bibitem [{\citenamefont {Muranaka}\ \emph {et~al.}(2008)\citenamefont
  {Muranaka}, \citenamefont {Kikuchi}, \citenamefont {Yoshizawa}, \citenamefont
  {Shirakawa},\ and\ \citenamefont
  {Akimitsu}}]{muranaka_superconductivity_2008}%
  \BibitemOpen
  \bibfield  {author} {\bibinfo {author} {\bibfnamefont {T.}~\bibnamefont
  {Muranaka}}, \bibinfo {author} {\bibfnamefont {Y.}~\bibnamefont {Kikuchi}},
  \bibinfo {author} {\bibfnamefont {T.}~\bibnamefont {Yoshizawa}}, \bibinfo
  {author} {\bibfnamefont {N.}~\bibnamefont {Shirakawa}}, \ and\ \bibinfo
  {author} {\bibfnamefont {J.}~\bibnamefont {Akimitsu}},\ }\href {\doibase
  10.1088/1468-6996/9/4/044204} {\bibfield  {journal} {\bibinfo  {journal}
  {Sci. Technol. Adv. Mater.}\ }\textbf {\bibinfo {volume} {9}},\ \bibinfo
  {pages} {044204} (\bibinfo {year} {2008})}\BibitemShut {NoStop}%
\bibitem [{\citenamefont {Militzer}\ and\ \citenamefont
  {Wilson}(2010)}]{militzer2010_prl-ice_cmcm}%
  \BibitemOpen
  \bibfield  {author} {\bibinfo {author} {\bibfnamefont {B.}~\bibnamefont
  {Militzer}}\ and\ \bibinfo {author} {\bibfnamefont {H.~F.}\ \bibnamefont
  {Wilson}},\ }\href@noop {} {\bibfield  {journal} {\bibinfo  {journal} {Phys.
  Rev. Lett.}\ }\textbf {\bibinfo {volume} {105}},\ \bibinfo {pages} {195701}
  (\bibinfo {year} {2010})}\BibitemShut {NoStop}%
\bibitem [{\citenamefont {McMahon}(2011)}]{mcmahon2011_rPRB_Ice-tera}%
  \BibitemOpen
  \bibfield  {author} {\bibinfo {author} {\bibfnamefont {J.~M.}\ \bibnamefont
  {McMahon}},\ }\href@noop {} {\bibfield  {journal} {\bibinfo  {journal} {Phys.
  Rev. B}\ }\textbf {\bibinfo {volume} {84}},\ \bibinfo {pages} {220104}
  (\bibinfo {year} {2011})}\BibitemShut {NoStop}%
\bibitem [{\citenamefont {Zhang}\ \emph {et~al.}(2013)\citenamefont {Zhang},
  \citenamefont {Wilson}, \citenamefont {Driver},\ and\ \citenamefont
  {Militzer}}]{H4O_coreplanet_PRB2013}%
  \BibitemOpen
  \bibfield  {author} {\bibinfo {author} {\bibfnamefont {S.}~\bibnamefont
  {Zhang}}, \bibinfo {author} {\bibfnamefont {H.~F.}\ \bibnamefont {Wilson}},
  \bibinfo {author} {\bibfnamefont {K.~P.}\ \bibnamefont {Driver}}, \ and\
  \bibinfo {author} {\bibfnamefont {B.}~\bibnamefont {Militzer}},\ }\href
  {\doibase 10.1103/PhysRevB.87.024112} {\bibfield  {journal} {\bibinfo
  {journal} {Phys. Rev. B}\ }\textbf {\bibinfo {volume} {87}},\ \bibinfo
  {pages} {024112} (\bibinfo {year} {2013})}\BibitemShut {NoStop}%
\bibitem [{\citenamefont {Pickard}\ \emph {et~al.}(2013)\citenamefont
  {Pickard}, \citenamefont {Martinez-Canales},\ and\ \citenamefont
  {Needs}}]{H2O2_Pickard-Needs_PRL2013}%
  \BibitemOpen
  \bibfield  {author} {\bibinfo {author} {\bibfnamefont {C.~J.}\ \bibnamefont
  {Pickard}}, \bibinfo {author} {\bibfnamefont {M.}~\bibnamefont
  {Martinez-Canales}}, \ and\ \bibinfo {author} {\bibfnamefont {R.~J.}\
  \bibnamefont {Needs}},\ }\href {\doibase 10.1103/PhysRevLett.110.245701}
  {\bibfield  {journal} {\bibinfo  {journal} {Phys. Rev. Lett.}\ }\textbf
  {\bibinfo {volume} {110}},\ \bibinfo {pages} {245701} (\bibinfo {year}
  {2013})}\BibitemShut {NoStop}%
\bibitem [{\citenamefont {Ma}\ \emph {et~al.}(2007)\citenamefont {Ma},
  \citenamefont {Oganov},\ and\ \citenamefont
  {Glass}}]{Ma-Oganov_oxygen_PRB2007}%
  \BibitemOpen
  \bibfield  {author} {\bibinfo {author} {\bibfnamefont {Y.}~\bibnamefont
  {Ma}}, \bibinfo {author} {\bibfnamefont {A.~R.}\ \bibnamefont {Oganov}}, \
  and\ \bibinfo {author} {\bibfnamefont {C.~W.}\ \bibnamefont {Glass}},\ }\href
  {\doibase 10.1103/PhysRevB.76.064101} {\bibfield  {journal} {\bibinfo
  {journal} {Phys. Rev. B}\ }\textbf {\bibinfo {volume} {76}},\ \bibinfo
  {pages} {064101} (\bibinfo {year} {2007})}\BibitemShut {NoStop}%
\bibitem [{\citenamefont {Ochoa-Calle}\ \emph {et~al.}(2015)\citenamefont
  {Ochoa-Calle}, \citenamefont {Zicovich-Wilson},\ and\ \citenamefont
  {Ram\'{\i}rez-Sol\'{\i}s}}]{Ochoa-calle_oxygen-hibrids_PRB2015}%
  \BibitemOpen
  \bibfield  {author} {\bibinfo {author} {\bibfnamefont {A.~J.}\ \bibnamefont
  {Ochoa-Calle}}, \bibinfo {author} {\bibfnamefont {C.~M.}\ \bibnamefont
  {Zicovich-Wilson}}, \ and\ \bibinfo {author} {\bibfnamefont {A.}~\bibnamefont
  {Ram\'{\i}rez-Sol\'{\i}s}},\ }\href {\doibase 10.1103/PhysRevB.92.085148}
  {\bibfield  {journal} {\bibinfo  {journal} {Phys. Rev. B}\ }\textbf {\bibinfo
  {volume} {92}},\ \bibinfo {pages} {085148} (\bibinfo {year}
  {2015})}\BibitemShut {NoStop}%
\bibitem [{\citenamefont {Weck}\ \emph {et~al.}(2009)\citenamefont {Weck},
  \citenamefont {Desgreniers}, \citenamefont {Loubeyre},\ and\ \citenamefont
  {Mezouar}}]{Mezouar_c2m-phase-Oxyg_PRL2009}%
  \BibitemOpen
  \bibfield  {author} {\bibinfo {author} {\bibfnamefont {G.}~\bibnamefont
  {Weck}}, \bibinfo {author} {\bibfnamefont {S.}~\bibnamefont {Desgreniers}},
  \bibinfo {author} {\bibfnamefont {P.}~\bibnamefont {Loubeyre}}, \ and\
  \bibinfo {author} {\bibfnamefont {M.}~\bibnamefont {Mezouar}},\ }\href
  {\doibase 10.1103/PhysRevLett.102.255503} {\bibfield  {journal} {\bibinfo
  {journal} {Phys. Rev. Lett.}\ }\textbf {\bibinfo {volume} {102}},\ \bibinfo
  {pages} {255503} (\bibinfo {year} {2009})}\BibitemShut {NoStop}%
\bibitem [{\citenamefont {Goncharov}\ \emph {et~al.}(1999)\citenamefont
  {Goncharov}, \citenamefont {Struzhkin}, \citenamefont {Mao},\ and\
  \citenamefont {Hemley}}]{Goncharov_PRL_Raman1998}%
  \BibitemOpen
  \bibfield  {author} {\bibinfo {author} {\bibfnamefont {A.~F.}\ \bibnamefont
  {Goncharov}}, \bibinfo {author} {\bibfnamefont {V.~V.}\ \bibnamefont
  {Struzhkin}}, \bibinfo {author} {\bibfnamefont {H.-k.}\ \bibnamefont {Mao}},
  \ and\ \bibinfo {author} {\bibfnamefont {R.~J.}\ \bibnamefont {Hemley}},\
  }\href {\doibase 10.1103/PhysRevLett.83.1998} {\bibfield  {journal} {\bibinfo
   {journal} {Phys. Rev. Lett.}\ }\textbf {\bibinfo {volume} {83}},\ \bibinfo
  {pages} {1998} (\bibinfo {year} {1999})}\BibitemShut {NoStop}%
\bibitem [{\citenamefont {Bartels-Rausch}\ \emph {et~al.}(2012)\citenamefont
  {Bartels-Rausch}, \citenamefont {Bergeron}, \citenamefont {Cartwright},
  \citenamefont {Escribano}, \citenamefont {Finney}, \citenamefont {Grothe},
  \citenamefont {Guti{\'e}rrez}, \citenamefont {Haapala}, \citenamefont {Kuhs},
  \citenamefont {Pettersson} \emph {et~al.}}]{REVIEW_2012ice}%
  \BibitemOpen
  \bibfield  {author} {\bibinfo {author} {\bibfnamefont {T.}~\bibnamefont
  {Bartels-Rausch}}, \bibinfo {author} {\bibfnamefont {V.}~\bibnamefont
  {Bergeron}}, \bibinfo {author} {\bibfnamefont {J.~H.}\ \bibnamefont
  {Cartwright}}, \bibinfo {author} {\bibfnamefont {R.}~\bibnamefont
  {Escribano}}, \bibinfo {author} {\bibfnamefont {J.~L.}\ \bibnamefont
  {Finney}}, \bibinfo {author} {\bibfnamefont {H.}~\bibnamefont {Grothe}},
  \bibinfo {author} {\bibfnamefont {P.~J.}\ \bibnamefont {Guti{\'e}rrez}},
  \bibinfo {author} {\bibfnamefont {J.}~\bibnamefont {Haapala}}, \bibinfo
  {author} {\bibfnamefont {W.~F.}\ \bibnamefont {Kuhs}}, \bibinfo {author}
  {\bibfnamefont {J.~B.}\ \bibnamefont {Pettersson}},  \emph {et~al.},\
  }\href@noop {} {\bibfield  {journal} {\bibinfo  {journal} {Rev. Mod. Phys.}\
  }\textbf {\bibinfo {volume} {84}},\ \bibinfo {pages} {885} (\bibinfo {year}
  {2012})}\BibitemShut {NoStop}%
\bibitem [{\citenamefont {Bjerrum}(1952)}]{bjerrum1952structure}%
  \BibitemOpen
  \bibfield  {author} {\bibinfo {author} {\bibfnamefont {N.}~\bibnamefont
  {Bjerrum}},\ }\href@noop {} {\bibfield  {journal} {\bibinfo  {journal}
  {Science}\ }\textbf {\bibinfo {volume} {115}},\ \bibinfo {pages} {385}
  (\bibinfo {year} {1952})}\BibitemShut {NoStop}%
\bibitem [{\citenamefont {Kamb}(1964)}]{kamb1964ice}%
  \BibitemOpen
  \bibfield  {author} {\bibinfo {author} {\bibfnamefont {B.}~\bibnamefont
  {Kamb}},\ }\href@noop {} {\bibfield  {journal} {\bibinfo  {journal} {Acta
  Crystallogr.}\ }\textbf {\bibinfo {volume} {17}},\ \bibinfo {pages} {1437}
  (\bibinfo {year} {1964})}\BibitemShut {NoStop}%
\bibitem [{\citenamefont {Londono}\ \emph {et~al.}(1993)\citenamefont
  {Londono}, \citenamefont {Kuhs},\ and\ \citenamefont
  {Finney}}]{londono1993neutron}%
  \BibitemOpen
  \bibfield  {author} {\bibinfo {author} {\bibfnamefont {J.}~\bibnamefont
  {Londono}}, \bibinfo {author} {\bibfnamefont {W.}~\bibnamefont {Kuhs}}, \
  and\ \bibinfo {author} {\bibfnamefont {J.}~\bibnamefont {Finney}},\
  }\href@noop {} {\bibfield  {journal} {\bibinfo  {journal} {J. Chem. Phys.}\
  }\textbf {\bibinfo {volume} {98}},\ \bibinfo {pages} {4878} (\bibinfo {year}
  {1993})}\BibitemShut {NoStop}%
\bibitem [{\citenamefont {Whalley}\ \emph {et~al.}(1968)\citenamefont
  {Whalley}, \citenamefont {Heath},\ and\ \citenamefont
  {Davidson}}]{whalley1968ice}%
  \BibitemOpen
  \bibfield  {author} {\bibinfo {author} {\bibfnamefont {E.}~\bibnamefont
  {Whalley}}, \bibinfo {author} {\bibfnamefont {J.}~\bibnamefont {Heath}}, \
  and\ \bibinfo {author} {\bibfnamefont {D.}~\bibnamefont {Davidson}},\
  }\href@noop {} {\bibfield  {journal} {\bibinfo  {journal} {J. Chem. Phys.}\
  }\textbf {\bibinfo {volume} {48}},\ \bibinfo {pages} {2362} (\bibinfo {year}
  {1968})}\BibitemShut {NoStop}%
\bibitem [{\citenamefont {Kuhs}\ \emph {et~al.}(1984)\citenamefont {Kuhs},
  \citenamefont {Finney}, \citenamefont {Vettier},\ and\ \citenamefont
  {Bliss}}]{kuhs1984structure}%
  \BibitemOpen
  \bibfield  {author} {\bibinfo {author} {\bibfnamefont {W.}~\bibnamefont
  {Kuhs}}, \bibinfo {author} {\bibfnamefont {J.}~\bibnamefont {Finney}},
  \bibinfo {author} {\bibfnamefont {C.}~\bibnamefont {Vettier}}, \ and\
  \bibinfo {author} {\bibfnamefont {D.}~\bibnamefont {Bliss}},\ }\href@noop {}
  {\bibfield  {journal} {\bibinfo  {journal} {J. Chem. Phys.}\ }\textbf
  {\bibinfo {volume} {81}},\ \bibinfo {pages} {3612} (\bibinfo {year}
  {1984})}\BibitemShut {NoStop}%
\bibitem [{\citenamefont {Kamb}\ \emph {et~al.}(1967)\citenamefont {Kamb},
  \citenamefont {Prakash},\ and\ \citenamefont {Knobler}}]{kamb1967structure}%
  \BibitemOpen
  \bibfield  {author} {\bibinfo {author} {\bibfnamefont {B.}~\bibnamefont
  {Kamb}}, \bibinfo {author} {\bibfnamefont {A.}~\bibnamefont {Prakash}}, \
  and\ \bibinfo {author} {\bibfnamefont {C.}~\bibnamefont {Knobler}},\
  }\href@noop {} {\bibfield  {journal} {\bibinfo  {journal} {Acta
  Crystallogr.}\ }\textbf {\bibinfo {volume} {22}},\ \bibinfo {pages} {706}
  (\bibinfo {year} {1967})}\BibitemShut {NoStop}%
\bibitem [{\citenamefont {Salzmann}\ \emph {et~al.}(2006)\citenamefont
  {Salzmann}, \citenamefont {Radaelli}, \citenamefont {Hallbrucker},
  \citenamefont {Mayer},\ and\ \citenamefont
  {Finney}}]{salzmann2006preparation}%
  \BibitemOpen
  \bibfield  {author} {\bibinfo {author} {\bibfnamefont {C.~G.}\ \bibnamefont
  {Salzmann}}, \bibinfo {author} {\bibfnamefont {P.~G.}\ \bibnamefont
  {Radaelli}}, \bibinfo {author} {\bibfnamefont {A.}~\bibnamefont
  {Hallbrucker}}, \bibinfo {author} {\bibfnamefont {E.}~\bibnamefont {Mayer}},
  \ and\ \bibinfo {author} {\bibfnamefont {J.~L.}\ \bibnamefont {Finney}},\
  }\href@noop {} {\bibfield  {journal} {\bibinfo  {journal} {Science}\ }\textbf
  {\bibinfo {volume} {311}},\ \bibinfo {pages} {1758} (\bibinfo {year}
  {2006})}\BibitemShut {NoStop}%
\bibitem [{\citenamefont {Jorgensen}\ and\ \citenamefont
  {Worlton}(1985)}]{jorgensen1985disordered}%
  \BibitemOpen
  \bibfield  {author} {\bibinfo {author} {\bibfnamefont {J.~D.}\ \bibnamefont
  {Jorgensen}}\ and\ \bibinfo {author} {\bibfnamefont {T.~G.}\ \bibnamefont
  {Worlton}},\ }\href@noop {} {\bibfield  {journal} {\bibinfo  {journal} {J.
  Chem. Phys.}\ }\textbf {\bibinfo {volume} {83}},\ \bibinfo {pages} {329}
  (\bibinfo {year} {1985})}\BibitemShut {NoStop}%
\bibitem [{\citenamefont {Besson}\ \emph {et~al.}(1994)\citenamefont {Besson},
  \citenamefont {Pruzan}, \citenamefont {Klotz}, \citenamefont {Hamel},
  \citenamefont {Silvi}, \citenamefont {Nelmes}, \citenamefont {Loveday},
  \citenamefont {Wilson},\ and\ \citenamefont {Hull}}]{besson1994variation}%
  \BibitemOpen
  \bibfield  {author} {\bibinfo {author} {\bibfnamefont {J.~M.}\ \bibnamefont
  {Besson}}, \bibinfo {author} {\bibfnamefont {P.}~\bibnamefont {Pruzan}},
  \bibinfo {author} {\bibfnamefont {S.}~\bibnamefont {Klotz}}, \bibinfo
  {author} {\bibfnamefont {G.}~\bibnamefont {Hamel}}, \bibinfo {author}
  {\bibfnamefont {B.}~\bibnamefont {Silvi}}, \bibinfo {author} {\bibfnamefont
  {R.~J.}\ \bibnamefont {Nelmes}}, \bibinfo {author} {\bibfnamefont {J.~S.}\
  \bibnamefont {Loveday}}, \bibinfo {author} {\bibfnamefont {R.~M.}\
  \bibnamefont {Wilson}}, \ and\ \bibinfo {author} {\bibfnamefont
  {S.}~\bibnamefont {Hull}},\ }\href {\doibase 10.1103/PhysRevB.49.12540}
  {\bibfield  {journal} {\bibinfo  {journal} {Phys. Rev. B}\ }\textbf {\bibinfo
  {volume} {49}},\ \bibinfo {pages} {12540} (\bibinfo {year}
  {1994})}\BibitemShut {NoStop}%
\bibitem [{\citenamefont {Hemley}\ \emph {et~al.}(1987)\citenamefont {Hemley},
  \citenamefont {Jephcoat}, \citenamefont {Mao}, \citenamefont {Zha},
  \citenamefont {Finger},\ and\ \citenamefont {Cox}}]{hemley1987static}%
  \BibitemOpen
  \bibfield  {author} {\bibinfo {author} {\bibfnamefont {R.}~\bibnamefont
  {Hemley}}, \bibinfo {author} {\bibfnamefont {A.}~\bibnamefont {Jephcoat}},
  \bibinfo {author} {\bibfnamefont {H.}~\bibnamefont {Mao}}, \bibinfo {author}
  {\bibfnamefont {C.}~\bibnamefont {Zha}}, \bibinfo {author} {\bibfnamefont
  {L.}~\bibnamefont {Finger}}, \ and\ \bibinfo {author} {\bibfnamefont
  {D.}~\bibnamefont {Cox}},\ }\href@noop {} {\bibfield  {journal} {\bibinfo
  {journal} {Nature}\ }\textbf {\bibinfo {volume} {330}},\ \bibinfo {pages}
  {737} (\bibinfo {year} {1987})}\BibitemShut {NoStop}%
\bibitem [{\citenamefont {Goncharov}\ \emph {et~al.}(2005)\citenamefont
  {Goncharov}, \citenamefont {Goldman}, \citenamefont {Fried}, \citenamefont
  {Crowhurst}, \citenamefont {Kuo}, \citenamefont {Mundy},\ and\ \citenamefont
  {Zaug}}]{goncharov2005dynamic}%
  \BibitemOpen
  \bibfield  {author} {\bibinfo {author} {\bibfnamefont {A.~F.}\ \bibnamefont
  {Goncharov}}, \bibinfo {author} {\bibfnamefont {N.}~\bibnamefont {Goldman}},
  \bibinfo {author} {\bibfnamefont {L.~E.}\ \bibnamefont {Fried}}, \bibinfo
  {author} {\bibfnamefont {J.~C.}\ \bibnamefont {Crowhurst}}, \bibinfo {author}
  {\bibfnamefont {I.~F.}\ \bibnamefont {Kuo}}, \bibinfo {author} {\bibfnamefont
  {C.~J.}\ \bibnamefont {Mundy}}, \ and\ \bibinfo {author} {\bibfnamefont
  {J.~M.}\ \bibnamefont {Zaug}},\ }\href {\doibase
  10.1103/PhysRevLett.94.125508} {\bibfield  {journal} {\bibinfo  {journal}
  {Phys. Rev. Lett.}\ }\textbf {\bibinfo {volume} {94}},\ \bibinfo {pages}
  {125508} (\bibinfo {year} {2005})}\BibitemShut {NoStop}%
\bibitem [{\citenamefont {Vos}\ \emph {et~al.}(1993)\citenamefont {Vos},
  \citenamefont {Finger}, \citenamefont {Hemley},\ and\ \citenamefont
  {Mao}}]{PRL_Vos-H2_Clathrate1993}%
  \BibitemOpen
  \bibfield  {author} {\bibinfo {author} {\bibfnamefont {W.~L.}\ \bibnamefont
  {Vos}}, \bibinfo {author} {\bibfnamefont {L.~W.}\ \bibnamefont {Finger}},
  \bibinfo {author} {\bibfnamefont {R.~J.}\ \bibnamefont {Hemley}}, \ and\
  \bibinfo {author} {\bibfnamefont {H.-k.}\ \bibnamefont {Mao}},\ }\href
  {\doibase 10.1103/PhysRevLett.71.3150} {\bibfield  {journal} {\bibinfo
  {journal} {Phys. Rev. Lett.}\ }\textbf {\bibinfo {volume} {71}},\ \bibinfo
  {pages} {3150} (\bibinfo {year} {1993})}\BibitemShut {NoStop}%
\bibitem [{\citenamefont {Mishima}(1994)}]{mishima1994reversible}%
  \BibitemOpen
  \bibfield  {author} {\bibinfo {author} {\bibfnamefont {O.}~\bibnamefont
  {Mishima}},\ }\href@noop {} {\bibfield  {journal} {\bibinfo  {journal} {J.
  Chem. Phys}\ }\textbf {\bibinfo {volume} {100}},\ \bibinfo {pages} {5910}
  (\bibinfo {year} {1994})}\BibitemShut {NoStop}%
\bibitem [{\citenamefont {Loerting}\ \emph {et~al.}(2001)\citenamefont
  {Loerting}, \citenamefont {Salzmann}, \citenamefont {Kohl}, \citenamefont
  {Mayer},\ and\ \citenamefont {Hallbrucker}}]{loerting2001second}%
  \BibitemOpen
  \bibfield  {author} {\bibinfo {author} {\bibfnamefont {T.}~\bibnamefont
  {Loerting}}, \bibinfo {author} {\bibfnamefont {C.}~\bibnamefont {Salzmann}},
  \bibinfo {author} {\bibfnamefont {I.}~\bibnamefont {Kohl}}, \bibinfo {author}
  {\bibfnamefont {E.}~\bibnamefont {Mayer}}, \ and\ \bibinfo {author}
  {\bibfnamefont {A.}~\bibnamefont {Hallbrucker}},\ }\href@noop {} {\bibfield
  {journal} {\bibinfo  {journal} {Phys. Chem. Chem. Phys.}\ }\textbf {\bibinfo
  {volume} {3}},\ \bibinfo {pages} {5355} (\bibinfo {year} {2001})}\BibitemShut
  {NoStop}%
\bibitem [{\citenamefont {Smith}\ and\ \citenamefont
  {Kay}(1999)}]{smith1999existence}%
  \BibitemOpen
  \bibfield  {author} {\bibinfo {author} {\bibfnamefont {R.~S.}\ \bibnamefont
  {Smith}}\ and\ \bibinfo {author} {\bibfnamefont {B.~D.}\ \bibnamefont
  {Kay}},\ }\href@noop {} {\bibfield  {journal} {\bibinfo  {journal} {Nature}\
  }\textbf {\bibinfo {volume} {398}},\ \bibinfo {pages} {788} (\bibinfo {year}
  {1999})}\BibitemShut {NoStop}%
\bibitem [{\citenamefont {Angell}(2004)}]{angell2004amorphous}%
  \BibitemOpen
  \bibfield  {author} {\bibinfo {author} {\bibfnamefont {C.~A.}\ \bibnamefont
  {Angell}},\ }\href@noop {} {\bibfield  {journal} {\bibinfo  {journal} {Annu.
  Rev. Phys. Chem.}\ }\textbf {\bibinfo {volume} {55}},\ \bibinfo {pages} {559}
  (\bibinfo {year} {2004})}\BibitemShut {NoStop}%
\bibitem [{\citenamefont {Yen}\ \emph {et~al.}(2015)\citenamefont {Yen},
  \citenamefont {Chi}, \citenamefont {Berlie}, \citenamefont {Liu},\ and\
  \citenamefont {Goncharov}}]{yen2015dielectric}%
  \BibitemOpen
  \bibfield  {author} {\bibinfo {author} {\bibfnamefont {F.}~\bibnamefont
  {Yen}}, \bibinfo {author} {\bibfnamefont {Z.}~\bibnamefont {Chi}}, \bibinfo
  {author} {\bibfnamefont {A.}~\bibnamefont {Berlie}}, \bibinfo {author}
  {\bibfnamefont {X.}~\bibnamefont {Liu}}, \ and\ \bibinfo {author}
  {\bibfnamefont {A.~F.}\ \bibnamefont {Goncharov}},\ }\href@noop {} {\bibfield
   {journal} {\bibinfo  {journal} {J. Phys. Chem. C}\ }\textbf {\bibinfo
  {volume} {119}},\ \bibinfo {pages} {20618} (\bibinfo {year}
  {2015})}\BibitemShut {NoStop}%
\bibitem [{\citenamefont {Goncharov}\ \emph {et~al.}(1996)\citenamefont
  {Goncharov}, \citenamefont {Struzhkin}, \citenamefont {Somayazulu},
  \citenamefont {Hemley},\ and\ \citenamefont
  {Mao}}]{goncharov1996compression}%
  \BibitemOpen
  \bibfield  {author} {\bibinfo {author} {\bibfnamefont {A.}~\bibnamefont
  {Goncharov}}, \bibinfo {author} {\bibfnamefont {V.}~\bibnamefont
  {Struzhkin}}, \bibinfo {author} {\bibfnamefont {M.}~\bibnamefont
  {Somayazulu}}, \bibinfo {author} {\bibfnamefont {R.}~\bibnamefont {Hemley}},
  \ and\ \bibinfo {author} {\bibfnamefont {H.}~\bibnamefont {Mao}},\
  }\href@noop {} {\bibfield  {journal} {\bibinfo  {journal} {Science}\ }\textbf
  {\bibinfo {volume} {273}},\ \bibinfo {pages} {218} (\bibinfo {year}
  {1996})}\BibitemShut {NoStop}%
\bibitem [{\citenamefont {Bove}\ \emph {et~al.}(2015)\citenamefont {Bove},
  \citenamefont {Gaal}, \citenamefont {Raza}, \citenamefont {Ludl},
  \citenamefont {Klotz}, \citenamefont {Saitta}, \citenamefont {Goncharov},\
  and\ \citenamefont {Gillet}}]{bove2015effect}%
  \BibitemOpen
  \bibfield  {author} {\bibinfo {author} {\bibfnamefont {L.~E.}\ \bibnamefont
  {Bove}}, \bibinfo {author} {\bibfnamefont {R.}~\bibnamefont {Gaal}}, \bibinfo
  {author} {\bibfnamefont {Z.}~\bibnamefont {Raza}}, \bibinfo {author}
  {\bibfnamefont {A.-A.}\ \bibnamefont {Ludl}}, \bibinfo {author}
  {\bibfnamefont {S.}~\bibnamefont {Klotz}}, \bibinfo {author} {\bibfnamefont
  {A.~M.}\ \bibnamefont {Saitta}}, \bibinfo {author} {\bibfnamefont {A.~F.}\
  \bibnamefont {Goncharov}}, \ and\ \bibinfo {author} {\bibfnamefont
  {P.}~\bibnamefont {Gillet}},\ }\href@noop {} {\bibfield  {journal} {\bibinfo
  {journal} {Proceedings of the National Academy of Sciences}\ }\textbf
  {\bibinfo {volume} {112}},\ \bibinfo {pages} {8216} (\bibinfo {year}
  {2015})}\BibitemShut {NoStop}%
\bibitem [{\citenamefont {Perdew}\ \emph {et~al.}(1996)\citenamefont {Perdew},
  \citenamefont {Burke},\ and\ \citenamefont {Ernzerhof}}]{PBE96}%
  \BibitemOpen
  \bibfield  {author} {\bibinfo {author} {\bibfnamefont {J.~P.}\ \bibnamefont
  {Perdew}}, \bibinfo {author} {\bibfnamefont {K.}~\bibnamefont {Burke}}, \
  and\ \bibinfo {author} {\bibfnamefont {M.}~\bibnamefont {Ernzerhof}},\
  }\href@noop {} {\bibfield  {journal} {\bibinfo  {journal} {Phys. Rev. Lett.}\
  }\textbf {\bibinfo {volume} {77}},\ \bibinfo {pages} {3865} (\bibinfo {year}
  {1996})}\BibitemShut {NoStop}%
\bibitem [{\citenamefont {Kresse}\ and\ \citenamefont
  {Furthm\"{u}ller}(1996)}]{VASP_Kresse}%
  \BibitemOpen
  \bibfield  {author} {\bibinfo {author} {\bibfnamefont {G.}~\bibnamefont
  {Kresse}}\ and\ \bibinfo {author} {\bibfnamefont {J.}~\bibnamefont
  {Furthm\"{u}ller}},\ }\href@noop {} {\bibfield  {journal} {\bibinfo
  {journal} {Comput. Mat. Sci.}\ }\textbf {\bibinfo {volume} {6}},\ \bibinfo
  {pages} {15} (\bibinfo {year} {1996})}\BibitemShut {NoStop}%
\bibitem [{\citenamefont {Liu}\ \emph {et~al.}(2015)\citenamefont {Liu},
  \citenamefont {Yao},\ and\ \citenamefont {Klug}}]{He-H2O_ice_PRB2016}%
  \BibitemOpen
  \bibfield  {author} {\bibinfo {author} {\bibfnamefont {H.}~\bibnamefont
  {Liu}}, \bibinfo {author} {\bibfnamefont {Y.}~\bibnamefont {Yao}}, \ and\
  \bibinfo {author} {\bibfnamefont {D.~D.}\ \bibnamefont {Klug}},\ }\href
  {\doibase 10.1103/PhysRevB.91.014102} {\bibfield  {journal} {\bibinfo
  {journal} {Phys. Rev. B}\ }\textbf {\bibinfo {volume} {91}},\ \bibinfo
  {pages} {014102} (\bibinfo {year} {2015})}\BibitemShut {NoStop}%
\bibitem [{\citenamefont {Hermann}\ \emph {et~al.}(2013)\citenamefont
  {Hermann}, \citenamefont {Ashcroft},\ and\ \citenamefont
  {Hoffmann}}]{Hermann_H2O_ZPE_PRB_2013}%
  \BibitemOpen
  \bibfield  {author} {\bibinfo {author} {\bibfnamefont {A.}~\bibnamefont
  {Hermann}}, \bibinfo {author} {\bibfnamefont {N.~W.}\ \bibnamefont
  {Ashcroft}}, \ and\ \bibinfo {author} {\bibfnamefont {R.}~\bibnamefont
  {Hoffmann}},\ }\href {\doibase 10.1103/PhysRevB.88.214113} {\bibfield
  {journal} {\bibinfo  {journal} {Phys. Rev. B}\ }\textbf {\bibinfo {volume}
  {88}},\ \bibinfo {pages} {214113} (\bibinfo {year} {2013})}\BibitemShut
  {NoStop}%
\bibitem [{\citenamefont {Benoit}\ \emph {et~al.}(1998)\citenamefont {Benoit},
  \citenamefont {Marx},\ and\ \citenamefont
  {Parrinello}}]{benoit1998tunnelling}%
  \BibitemOpen
  \bibfield  {author} {\bibinfo {author} {\bibfnamefont {M.}~\bibnamefont
  {Benoit}}, \bibinfo {author} {\bibfnamefont {D.}~\bibnamefont {Marx}}, \ and\
  \bibinfo {author} {\bibfnamefont {M.}~\bibnamefont {Parrinello}},\
  }\href@noop {} {\bibfield  {journal} {\bibinfo  {journal} {Nature}\ }\textbf
  {\bibinfo {volume} {392}},\ \bibinfo {pages} {258} (\bibinfo {year}
  {1998})}\BibitemShut {NoStop}%
\bibitem [{\citenamefont {Becke}\ and\ \citenamefont
  {Edgecombe}(1990)}]{Becke_ELF_JChPh1990}%
  \BibitemOpen
  \bibfield  {author} {\bibinfo {author} {\bibfnamefont {A.~D.}\ \bibnamefont
  {Becke}}\ and\ \bibinfo {author} {\bibfnamefont {K.~E.}\ \bibnamefont
  {Edgecombe}},\ }\href {\doibase http://dx.doi.org/10.1063/1.458517}
  {\bibfield  {journal} {\bibinfo  {journal} {J. Chem. Phys.}\ }\textbf
  {\bibinfo {volume} {92}},\ \bibinfo {pages} {5397} (\bibinfo {year}
  {1990})}\BibitemShut {NoStop}%
\bibitem [{\citenamefont {Glawe}\ \emph {et~al.}(2016)\citenamefont {Glawe},
  \citenamefont {Sanna}, \citenamefont {Gross},\ and\ \citenamefont
  {Marques}}]{Henning-Sanna-Marques_pettifor}%
  \BibitemOpen
  \bibfield  {author} {\bibinfo {author} {\bibfnamefont {H.}~\bibnamefont
  {Glawe}}, \bibinfo {author} {\bibfnamefont {A.}~\bibnamefont {Sanna}},
  \bibinfo {author} {\bibfnamefont {E.~K.~U.}\ \bibnamefont {Gross}}, \ and\
  \bibinfo {author} {\bibfnamefont {M.~A.~L.}\ \bibnamefont {Marques}},\ }\href
  {http://stacks.iop.org/1367-2630/18/i=9/a=093011} {\bibfield  {journal}
  {\bibinfo  {journal} {New Journal of Physics}\ }\textbf {\bibinfo {volume}
  {18}},\ \bibinfo {pages} {093011} (\bibinfo {year} {2016})}\BibitemShut
  {NoStop}%
\bibitem [{\citenamefont {Blase}(2011)}]{blase_superconductivity_2011}%
  \BibitemOpen
  \bibfield  {author} {\bibinfo {author} {\bibfnamefont {X.}~\bibnamefont
  {Blase}},\ }\href {\doibase 10.1016/j.crhy.2011.03.002} {\bibfield  {journal}
  {\bibinfo  {journal} {C. R. Phys.}\ }\textbf {\bibinfo {volume} {12}},\
  \bibinfo {pages} {584} (\bibinfo {year} {2011})}\BibitemShut {NoStop}%
\bibitem [{\citenamefont {Eliashberg}(1960)}]{Eliashberg}%
  \BibitemOpen
  \bibfield  {author} {\bibinfo {author} {\bibfnamefont {G.}~\bibnamefont
  {Eliashberg}},\ }\href@noop {} {\bibfield  {journal} {\bibinfo  {journal}
  {Teor. Fiz.}\ }\textbf {\bibinfo {volume} {38}} (\bibinfo {year} {1960})},\
  \bibinfo {note} {[Sov. Phys. FETP {\bf 11}, 696 (1960)]}\BibitemShut
  {NoStop}%
\bibitem [{\citenamefont {Allen}\ and\ \citenamefont {Mitrovi{\'
  c}}(1983)}]{AllenMitrovic1983}%
  \BibitemOpen
  \bibfield  {author} {\bibinfo {author} {\bibfnamefont {P.~B.}\ \bibnamefont
  {Allen}}\ and\ \bibinfo {author} {\bibfnamefont {B.}~\bibnamefont {Mitrovi{\'
  c}}},\ }\href {\doibase http://dx.doi.org/10.1016/S0081-1947(08)60665-7}
  {\emph {\bibinfo {title} {Theory of Superconducting Tc}}},\ \bibinfo {series}
  {Solid State Physics}, Vol.~\bibinfo {volume} {37}\ (\bibinfo  {publisher}
  {Academic Press},\ \bibinfo {year} {1983})\ pp.\ \bibinfo {pages} {1 --
  92}\BibitemShut {NoStop}%
\bibitem [{\citenamefont {Kohn}(1959)}]{KohnAnomaly}%
  \BibitemOpen
  \bibfield  {author} {\bibinfo {author} {\bibfnamefont {W.}~\bibnamefont
  {Kohn}},\ }\href {\doibase 10.1103/PhysRevLett.2.393} {\bibfield  {journal}
  {\bibinfo  {journal} {Phys. Rev. Lett.}\ }\textbf {\bibinfo {volume} {2}},\
  \bibinfo {pages} {393} (\bibinfo {year} {1959})}\BibitemShut {NoStop}%
\bibitem [{Note1()}]{Note1}%
  \BibitemOpen
  \bibinfo {note} {The phonon spectra and the electron-phonon matrix elements
  were obtained employing density-functional perturbation theory~\cite
  {Baroni_1987a,2n+1_Gonzepaper}, as implemented in the plane-wave based code
  {\protect \sc abinit}~\cite {gonze_abinit_2009}. For the electron-phonon the
  following $k$ and $q$-meshes were used for the different supercells: 25\%
  doping, $k=16\times 16\times 16$, and $q=8\times 8\times 8$; 12.5\% doping,
  $k=8\times 8\times 8$ and $q=4\times 4\times 4$; 6.13\% doping, $k=4\times
  4\times 4$ and $q=4\times 4\times 4$; 4.17\% doping, $k=2\times 2\times 2$
  and $q=2\times 2\times 2$.}\BibitemShut {Stop}%
\bibitem [{\citenamefont {Carbotte}(1990)}]{Carbotte_RMP1990}%
  \BibitemOpen
  \bibfield  {author} {\bibinfo {author} {\bibfnamefont {J.~P.}\ \bibnamefont
  {Carbotte}},\ }\href {\doibase 10.1103/RevModPhys.62.1027} {\bibfield
  {journal} {\bibinfo  {journal} {Rev. Mod. Phys.}\ }\textbf {\bibinfo {volume}
  {62}},\ \bibinfo {pages} {1027} (\bibinfo {year} {1990})}\BibitemShut
  {NoStop}%
\bibitem [{\citenamefont {Allen}\ and\ \citenamefont
  {Dynes}(1975)}]{AllenDynes_PRB1975}%
  \BibitemOpen
  \bibfield  {author} {\bibinfo {author} {\bibfnamefont {P.~B.}\ \bibnamefont
  {Allen}}\ and\ \bibinfo {author} {\bibfnamefont {R.~C.}\ \bibnamefont
  {Dynes}},\ }\href {\doibase 10.1103/PhysRevB.12.905} {\bibfield  {journal}
  {\bibinfo  {journal} {Phys. Rev. B}\ }\textbf {\bibinfo {volume} {12}},\
  \bibinfo {pages} {905} (\bibinfo {year} {1975})}\BibitemShut {NoStop}%
\bibitem [{\citenamefont {Smirnov}\ and\ \citenamefont
  {Stegailov}(2013)}]{Grigory_JPL_clathrate-2013}%
  \BibitemOpen
  \bibfield  {author} {\bibinfo {author} {\bibfnamefont {G.~S.}\ \bibnamefont
  {Smirnov}}\ and\ \bibinfo {author} {\bibfnamefont {V.~V.}\ \bibnamefont
  {Stegailov}},\ }\href {\doibase 10.1021/jz401669d} {\bibfield  {journal}
  {\bibinfo  {journal} {The Journal of Physical Chemistry Letters}\ }\textbf
  {\bibinfo {volume} {4}},\ \bibinfo {pages} {3560} (\bibinfo {year}
  {2013})}\BibitemShut {NoStop}%
\bibitem [{\citenamefont {Qian}\ \emph {et~al.}(2014)\citenamefont {Qian},
  \citenamefont {Lyakhov}, \citenamefont {Zhu}, \citenamefont {Oganov},\ and\
  \citenamefont {Dong}}]{Quian_SR_hydrogen-clathrates2013}%
  \BibitemOpen
  \bibfield  {author} {\bibinfo {author} {\bibfnamefont {G.-R.}\ \bibnamefont
  {Qian}}, \bibinfo {author} {\bibfnamefont {A.~O.}\ \bibnamefont {Lyakhov}},
  \bibinfo {author} {\bibfnamefont {Q.}~\bibnamefont {Zhu}}, \bibinfo {author}
  {\bibfnamefont {A.~R.}\ \bibnamefont {Oganov}}, \ and\ \bibinfo {author}
  {\bibfnamefont {X.}~\bibnamefont {Dong}},\ }\href@noop {} {\bibfield
  {journal} {\bibinfo  {journal} {Scientific reports}\ }\textbf {\bibinfo
  {volume} {4}} (\bibinfo {year} {2014})}\BibitemShut {NoStop}%
\bibitem [{\citenamefont {Strobel}\ \emph
  {et~al.}(2011{\natexlab{b}})\citenamefont {Strobel}, \citenamefont
  {Somayazulu},\ and\ \citenamefont {Hemley}}]{Strobel_H2_H2O_2011}%
  \BibitemOpen
  \bibfield  {author} {\bibinfo {author} {\bibfnamefont {T.~A.}\ \bibnamefont
  {Strobel}}, \bibinfo {author} {\bibfnamefont {M.}~\bibnamefont {Somayazulu}},
  \ and\ \bibinfo {author} {\bibfnamefont {R.~J.}\ \bibnamefont {Hemley}},\
  }\href {\doibase 10.1021/jp1122536} {\bibfield  {journal} {\bibinfo
  {journal} {The Journal of Physical Chemistry C}\ }\textbf {\bibinfo {volume}
  {115}},\ \bibinfo {pages} {4898} (\bibinfo {year}
  {2011}{\natexlab{b}})}\BibitemShut {NoStop}%
\bibitem [{\citenamefont {Baroni}\ \emph {et~al.}(1987)\citenamefont {Baroni},
  \citenamefont {Giannozzi},\ and\ \citenamefont {Testa}}]{Baroni_1987a}%
  \BibitemOpen
  \bibfield  {author} {\bibinfo {author} {\bibfnamefont {S.}~\bibnamefont
  {Baroni}}, \bibinfo {author} {\bibfnamefont {P.}~\bibnamefont {Giannozzi}}, \
  and\ \bibinfo {author} {\bibfnamefont {A.}~\bibnamefont {Testa}},\ }\href
  {\doibase 10.1103/PhysRevLett.58.1861} {\bibfield  {journal} {\bibinfo
  {journal} {Phys. Rev. Lett.}\ }\textbf {\bibinfo {volume} {58}},\ \bibinfo
  {pages} {1861} (\bibinfo {year} {1987})}\BibitemShut {NoStop}%
\bibitem [{\citenamefont {Gonze}\ and\ \citenamefont
  {Vigneron}(1989)}]{2n+1_Gonzepaper}%
  \BibitemOpen
  \bibfield  {author} {\bibinfo {author} {\bibfnamefont {X.}~\bibnamefont
  {Gonze}}\ and\ \bibinfo {author} {\bibfnamefont {J.-P.}\ \bibnamefont
  {Vigneron}},\ }\href {\doibase 10.1103/PhysRevB.39.13120} {\bibfield
  {journal} {\bibinfo  {journal} {Phys. Rev. B}\ }\textbf {\bibinfo {volume}
  {39}},\ \bibinfo {pages} {13120} (\bibinfo {year} {1989})}\BibitemShut
  {NoStop}%
\bibitem [{\citenamefont {Gonze}\ \emph {et~al.}(2009)\citenamefont {Gonze},
  \citenamefont {Amadon}, \citenamefont {Anglade}, \citenamefont {Beuken},
  \citenamefont {Bottin}, \citenamefont {Boulanger}, \citenamefont {Bruneval},
  \citenamefont {Caliste}, \citenamefont {Caracas}, \citenamefont
  {C\^{o}t\'{e}}, \citenamefont {Deutsch}, \citenamefont {Genovese},
  \citenamefont {Ghosez}, \citenamefont {Giantomassi}, \citenamefont
  {Goedecker}, \citenamefont {Hamann}, \citenamefont {Hermet}, \citenamefont
  {Jollet}, \citenamefont {Jomard}, \citenamefont {Leroux}, \citenamefont
  {Mancini}, \citenamefont {Mazevet}, \citenamefont {Oliveira}, \citenamefont
  {Onida}, \citenamefont {Pouillon}, \citenamefont {Rangel}, \citenamefont
  {Rignanese}, \citenamefont {Sangalli}, \citenamefont {Shaltaf}, \citenamefont
  {Torrent}, \citenamefont {Verstraete}, \citenamefont {Zerah},\ and\
  \citenamefont {Zwanziger}}]{gonze_abinit_2009}%
  \BibitemOpen
  \bibfield  {author} {\bibinfo {author} {\bibfnamefont {X.}~\bibnamefont
  {Gonze}}, \bibinfo {author} {\bibfnamefont {B.}~\bibnamefont {Amadon}},
  \bibinfo {author} {\bibfnamefont {P.}~\bibnamefont {Anglade}}, \bibinfo
  {author} {\bibfnamefont {J.}~\bibnamefont {Beuken}}, \bibinfo {author}
  {\bibfnamefont {F.}~\bibnamefont {Bottin}}, \bibinfo {author} {\bibfnamefont
  {P.}~\bibnamefont {Boulanger}}, \bibinfo {author} {\bibfnamefont
  {F.}~\bibnamefont {Bruneval}}, \bibinfo {author} {\bibfnamefont
  {D.}~\bibnamefont {Caliste}}, \bibinfo {author} {\bibfnamefont
  {R.}~\bibnamefont {Caracas}}, \bibinfo {author} {\bibfnamefont
  {M.}~\bibnamefont {C\^{o}t\'{e}}}, \bibinfo {author} {\bibfnamefont
  {T.}~\bibnamefont {Deutsch}}, \bibinfo {author} {\bibfnamefont
  {L.}~\bibnamefont {Genovese}}, \bibinfo {author} {\bibfnamefont
  {P.}~\bibnamefont {Ghosez}}, \bibinfo {author} {\bibfnamefont
  {M.}~\bibnamefont {Giantomassi}}, \bibinfo {author} {\bibfnamefont
  {S.}~\bibnamefont {Goedecker}}, \bibinfo {author} {\bibfnamefont
  {D.}~\bibnamefont {Hamann}}, \bibinfo {author} {\bibfnamefont
  {P.}~\bibnamefont {Hermet}}, \bibinfo {author} {\bibfnamefont
  {F.}~\bibnamefont {Jollet}}, \bibinfo {author} {\bibfnamefont
  {G.}~\bibnamefont {Jomard}}, \bibinfo {author} {\bibfnamefont
  {S.}~\bibnamefont {Leroux}}, \bibinfo {author} {\bibfnamefont
  {M.}~\bibnamefont {Mancini}}, \bibinfo {author} {\bibfnamefont
  {S.}~\bibnamefont {Mazevet}}, \bibinfo {author} {\bibfnamefont
  {M.}~\bibnamefont {Oliveira}}, \bibinfo {author} {\bibfnamefont
  {G.}~\bibnamefont {Onida}}, \bibinfo {author} {\bibfnamefont
  {Y.}~\bibnamefont {Pouillon}}, \bibinfo {author} {\bibfnamefont
  {T.}~\bibnamefont {Rangel}}, \bibinfo {author} {\bibfnamefont
  {G.}~\bibnamefont {Rignanese}}, \bibinfo {author} {\bibfnamefont
  {D.}~\bibnamefont {Sangalli}}, \bibinfo {author} {\bibfnamefont
  {R.}~\bibnamefont {Shaltaf}}, \bibinfo {author} {\bibfnamefont
  {M.}~\bibnamefont {Torrent}}, \bibinfo {author} {\bibfnamefont
  {M.}~\bibnamefont {Verstraete}}, \bibinfo {author} {\bibfnamefont
  {G.}~\bibnamefont {Zerah}}, \ and\ \bibinfo {author} {\bibfnamefont
  {J.}~\bibnamefont {Zwanziger}},\ }\href@noop {} {\bibfield  {journal}
  {\bibinfo  {journal} {Comput. Phys. Commun.}\ }\textbf {\bibinfo {volume}
  {180}},\ \bibinfo {pages} {2582} (\bibinfo {year} {2009})}\BibitemShut
  {NoStop}%
\end{thebibliography}%

\onecolumngrid
\newpage

\begin{center}
\Huge Supplemental Material
\end{center}

\section*{Hole doped structure of ice at high pressure}

\begin{figure}[htb]
  \begin{center}
    \includegraphics[width=1.0\columnwidth,angle=0]{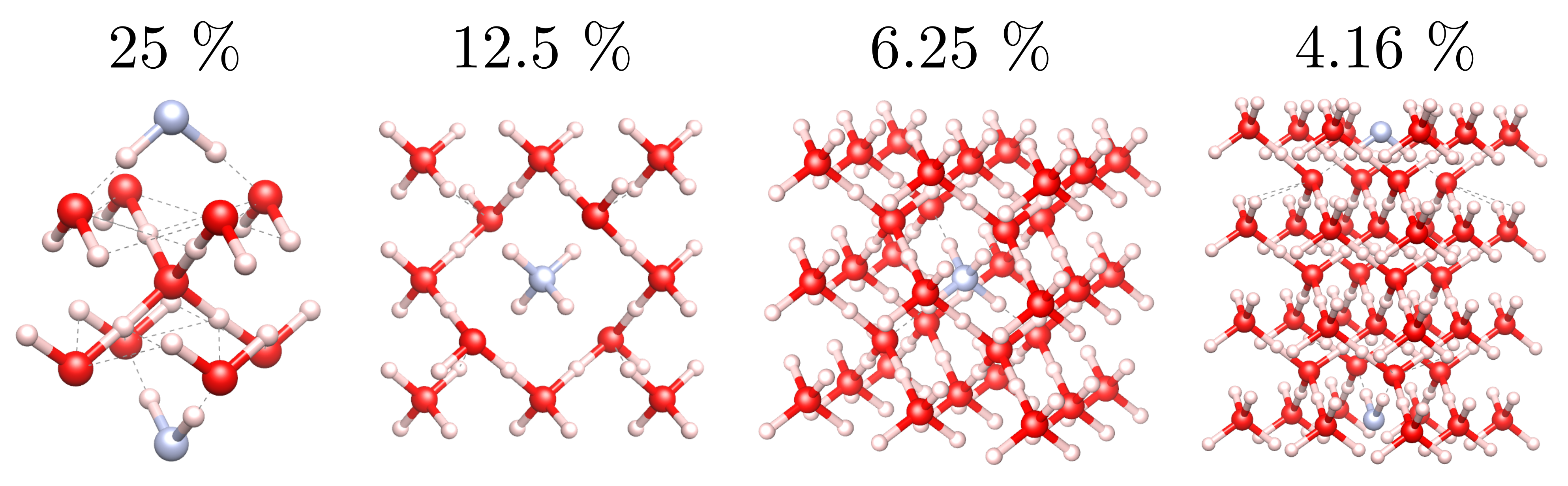}
    \caption{(Color online) Simulation supercells used at different doping levels. }
    \label{fig:N-doped_STRUC}
  \end{center}
\end{figure}

In order to simulate the hole doped ice crystal under pressure we substitute oxygen 
for nitrogen in ratios of  25\%, 12\%, 6\% and 4\% in the supercells model show in 
Fig.~\ref{fig:N-doped_STRUC}.  
Structural relaxation then were carried out for the supercells at 150\,GPa of pressure. 
We found that low doping (4--6\%) induces fairly small modifications to the ice-X crystal structure, 
while larger doping levels have a stronger effect on the local environment, 
as reflected in a volume expansion. Still the ice-x structural motif is preserved for the 
substitution/doping studied in this work independently of the site symmetry. 

\section*{Stability and formation enthalpy H--O--N system}

\begin{figure*}[htb]
  \begin{center}
    \includegraphics[width=0.95\columnwidth,angle=0]{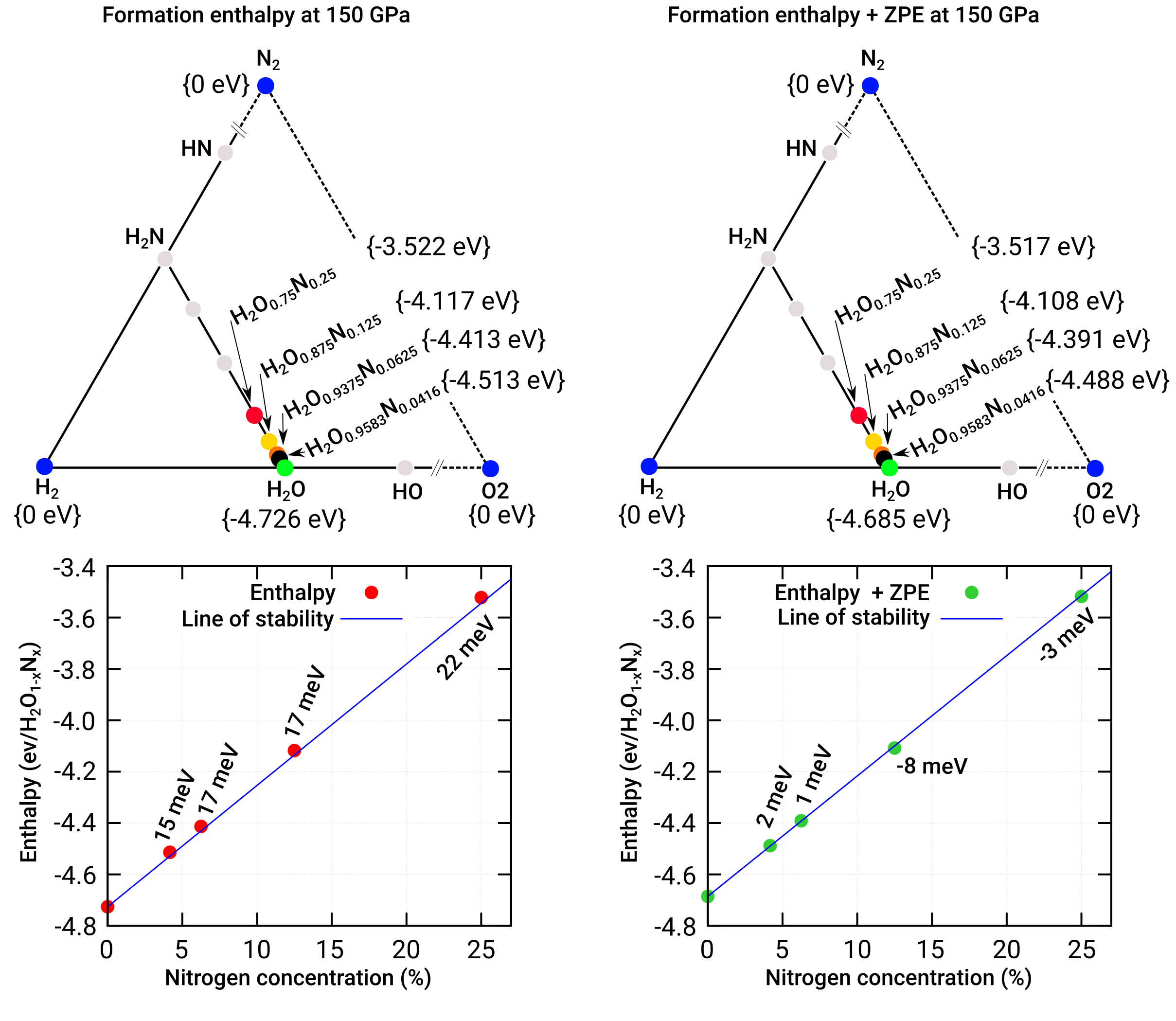}
    \caption{(Color online) Left panel shows the formation enthalpy
      for the doped structures at 150\,GPa. Right panel shows the
      free energy including the zero point energy correction.  Bottom
      panels represent the line of stability of the ternary phase
      diagram between H$_2$O and an hypothetical H$_2$N composition. }
    \label{fig:TernaryP-D}
  \end{center}
\end{figure*}

We studied the stability of the doped compositions with nitrogen by means of total energy DFT calculations.  
The partial phase diagram for the ternary phase (H--O--N) is shown in the top panels of Fig.~\ref{fig:TernaryP-D}.  
Solid blue dots represent the ground-state (experimentally know structures) elemental 
phases of H$_2$, O$_2$ and N$_2$ occurring at 150\,GPa.  
We find the phase X of ice stable as depicted by green circles.  
In the two bottom panels we plot the stability line between H$_2$O to an hypothetical H$_2$N composition respectively without (left) and with (right) zero point energy corrections (ZPE).
Without ZPE all doped structures (4\%, 6\%, 12\%, and 25\%) lie only $\sim$15\,meV per formula unit above the stability line, and the structures are further stabilized by ZPE: for low doping (4\% and 6\%) the structures are a mere 2\,meV per formula unit above the stability line (metastable compositions);
while for 12\% and 25\% of nitrogen we find the doped structures well below the stability line. 
In fact, within our computational accuracy, we can claim these stoichiometries as new stable 
points in the ternary phase diagram of H--O--N. 

\section*{Crystal structure for N-doped ice-X}

In Table~1 we give details of the supercell model of N-doped ice-X used for this work. 

\begin{table}
  \begin{center}
 \caption{Crystal structure parameters for the supercell created to dope the phase-X of ice (relaxations at indicated pressure).} \vspace{0.25cm}
\begin{tabular}{ c | c | c | c | c | c }
 Composition & Crystal    & Lattice parameters       & Element  & Site symmetry  & Atom coordinates           \\
  (Pressure) & space group    &          (\AA)           &     & Wyckoff position    & (internal coordinates)     \\ \hline   \hline
H$_2$O$_{0.75}$N$_{0.25}$ & $P42m$ &  $a=$ 2.52145   &   H   & $4n$ & 0.24756 0.24756 0.62448  \\
 (150\,GPa)               & (111)  &  $b=$ 2.52145   &   H   & $4n$ & 0.73386 0.73386 0.11316  \\
                          &        &  $c=$ 5.56810   &   O   & $2g$ & 0.00000 0.00000 0.73886  \\
                          &        &  $\alpha=$ 90.0 &   O   & $1b$ & 0.50000 0.50000 0.50000  \\
                          &        &  $\beta=$  90.0 &   N   & $1d$ & 0.50000 0.50000 0.00000  \\
                          &        &  $\gamma=$ 90.0 &       &      &   \\  \hline
H$_2$O$_{0.875}$N$_{0.125}$ & $P42m$ &  $a=$ 5.25065   &   H   & $8o$ & 0.87574 0.37395 0.74495  \\
 (150\,GPa)                 & (111)  &  $b=$ 5.25065   &   H   & $4n$ & 0.61458 0.61458 0.73580  \\
                            &        &  $c=$ 2.54502   &   H   & $4n$ & 0.87426 0.87426 0.75826  \\
                            &        &  $\alpha=$ 90.0 &   O   & $4n$ & 0.75602 0.75602 0.01405  \\
                            &        &  $\beta=$  90.0 &   O   & $2f$ & 0.50000 0.00000 0.50000  \\
                            &        &  $\gamma=$ 90.0 &   O   & $1c$ & 0.00000 0.00000 0.50000  \\ 
                            &        &                 &   N   & $1b$ & 0.50000 0.50000 0.50000  \\   \hline
H$_2$O$_{0.9375}$N$_{0.0625}$ & $P43m$ &  $a=$ 5.18627   &   H   & $12i$ & 0.37602 0.37602 0.12528  \\
 (150\,GPa)                   & (215)  &  $b=$ 5.18627   &   H   & $4e$  & 0.12670 0.12670 0.12670  \\
                              &        &  $c=$ 5.18627   &   H   & $12i$ & 0.12407 0.12407 0.62727  \\
                              &        &  $\alpha=$ 90.0 &   H   & $4e$  & 0.61579 0.61579 0.61579  \\
                              &        &  $\beta=$  90.0 &   O   & $4e$  & 0.25038 0.25038 0.25038  \\
                              &        &  $\gamma=$ 90.0 &   O   & $3c$  & 0.00000 0.50000 0.50000  \\                       
                              &        &                 &   O   & $4e$  & 0.75618 0.75618 0.75618  \\  
                              &        &                 &   O   & $3d$  & 0.50000 0.00000 0.00000  \\   
                              &        &                 &   O   & $1a$  & 0.00000 0.00000 0.00000  \\   
                              &        &                 &   N   & $1b$  & 0.50000 0.50000 0.50000  \\    \hline
H$_2$O$_{0.9583}$N$_{0.0416}$ & $P42m$ &  $a=$ 5.17922   &   H   & $8o$  & 0.87522 0.36997 0.74963  \\
 (150\,GPa)                   & (111)  &  $b=$ 5.17922   &   H   & $4n$  & 0.61393 0.61393 0.07818  \\
                              &        &  $c=$ 7.78908   &   H   & $4n$  & 0.87617 0.87617 0.08474  \\
                              &        &  $\alpha=$ 90.0 &   H   & $8o$  & 0.87473 0.37526 0.08288  \\
                              &        &  $\beta=$  90.0 &   H   & $4n$  & 0.37527 0.37527 0.41564  \\
                              &        &  $\gamma=$ 90.0 &   H   & $8o$  & 0.87591 0.37826 0.41577  \\                    
                              &        &                 &   H   & $4n$  & 0.87198 0.87198 0.41991  \\
                              &        &                 &   H   & $4n$  & 0.37459 0.37459 0.75185  \\  
                              &        &                 &   H   & $4n$  & 0.87614 0.87614 0.74795  \\  
                              &        &                 &   O   & $4n$  & 0.75689 0.75689 0.17137  \\   
                              &        &                 &   O   & $4m$  & 0.00000 0.50000 0.66554  \\   
                              &        &                 &   O   & $4n$  & 0.75142 0.75142 0.83303  \\   
                              &        &                 &   O   & $4n$  & 0.25202 0.25202 0.50031   \\   
                              &        &                 &   O   & $2e$  & 0.50000 0.00000 0.00000  \\   
                              &        &                 &   O   & $1a$  & 0.00000 0.00000 0.00000  \\   
                              &        &                 &   O   & $2h$  & 0.50000 0.50000 0.33125  \\   
                              &        &                 &   O   & $2g$  & 0.00000 0.00000 0.33678  \\   
                              &        &                 &   N   & $1d$  & 0.50000 0.50000 0.00000  \\    \hline
\end{tabular}
\end{center}
\end{table}

\end{document}